\documentclass[twocolumn,twocolappendix]{aastex631}

\usepackage{mathtools}

\newcommand{\gaiasource}{\texttt{gaia\_source}}
\newcommand{\nsstbo}{\texttt{nss\_two\_body\_orbit}}
\newcommand{\nssacc}{\texttt{nss\_acceleration\_astro}}

\newcommand{\ruwe}{\texttt{ruwe}}

\begin{document}

\title{A Fast, Analytic Empirical Model of the \emph{Gaia} Data Release 3 Astrometric Orbit Catalog Selection Function}

\author[0000-0002-6406-1924]{Casey Y. Lam}
\correspondingauthor{Casey Y. Lam} 
\email{clam@carnegiescience.edu}
\affiliation{Observatories of the Carnegie Institution for Science, Pasadena, CA 91101, USA}

\author[0000-0002-6871-1752]{Kareem El-Badry}
\affiliation{Department of Astronomy, California Institute of Technology, 1200 E. California Blvd., Pasadena, CA 91125, USA}

\author[0000-0002-4733-4994]{Joshua D. Simon}
\affiliation{Observatories of the Carnegie Institution for Science, Pasadena, CA 91101, USA}

\begin{abstract}

In June 2022, the \emph{Gaia} mission released a catalog of astrometric orbital solutions for 168,065 binary systems, by far the largest such catalog to date.
% Unlike previous binary stars catalogs, which were heterogeneous collections of orbits from different surveys and instruments, these orbits were derived using \emph{Gaia} data alone.
% Despite this homogeneity, 
The catalog's selection function is difficult to characterize because of choices made in its construction. % of the catalog.
Understanding the catalog's selection function is required to model and interpret its contents.
We use a combination of analytic and empirical prescriptions to construct a function that computes the probability that a binary with a given set of properties would have been published in the \emph{Gaia} Data Release 3 astrometric orbit catalog.
This is a complementary approach to the more accurate but significantly more computationally expensive approach of \cite{Kareem}.
We also construct a binary population synthesis model 
% based on \cite{Moe:2017} 
to validate our characterization of the selection function, finding good agreement with the actual \emph{Gaia} NSS catalog, with the exception of the orbital eccentricity distribution.
The NSS catalog suggests high-eccentricity orbits are relatively uncommon at intermediate periods $100 \lesssim P_{orb} \lesssim 1000$ days.
As an example application of the selection function, we estimate the \emph{Gaia} DR3 detection probabilities of the star + BH binaries Gaia BH1 and BH2, and find them to be 0.38 and  0.27, respectively.
Compared to the values obtained by detailed modeling in \cite{Kareem}, the probabilities are identical for BH1, and within a factor of 2 for BH2.
We also estimate the population of Sun-like star + BH binaries in the Galaxy to be $\sim 3000$ for $100 < P_{orb} < 400$ day, $< 800$ for $400 < P_{orb} < 1000$ day, and $< 12,000$ for $1000 < P_{orb} < 1500$ day.
\end{abstract}

% \keywords{Fill in later.}

\section{Introduction
\label{sec:Introduction}}

\emph{Gaia} is an astrometric space mission surveying over 2 billion stars in the Milky Way \citep{Prusti:2016}. 
Its main mission operated from July 2014 through December 2019, and its extended mission operated through January 2025.
The data products and catalogs generated from the photometry, astrometry, and spectroscopy from \emph{Gaia} are shared publicly in data releases that occur every several years.

In previous \emph{Gaia} data releases, all catalog parameters were produced using single-star models \citep{GaiaDR1:2016, GaiaDR2:2018, GaiaEDR3:2021}.
For the first time in Data Release 3 (DR3), a catalog of non-single stars (NSS) was produced \citep{GaiaDR3:2023, Arenou:2023}.
A wide variety of science involving binary and multiple stars has already been done with the NSS catalog since it was released in June 2022, including the first mass measurement of a star orbiting a detached black hole \citep{El-Badry:2023a,Chakrabarti:2023}; for a summary of other discoveries, see \cite{El-Badry:2024}.

Performing population studies of the stellar and compact remnant binaries found in \emph{Gaia} DR3 requires understanding the selection effects of the NSS catalog.
The \emph{Gaia} collaboration cautions that this is non-trivial to derive from their processing pipeline \citep{Halbwachs:2023}.
Furthermore, all \emph{Gaia} data releases so far, including DR3, only provide model parameters, and not the individual photometric, spectroscopic, or astrometric measurements; without the starting input, it is not possible to exactly reproduce the processing pipeline or perform injection and recovery tests.

In this paper, we perform an empirical forward model of the \emph{Gaia} NSS astrometric orbits catalog.
We focus on the solutions that have an astrometric orbit (as compared to those that only have a spectroscopic orbit) as our primary goal is to understand the selection function of stellar-mass black hole + star binaries, which have thus far only been found in the astrometric orbits catalog.
However, we emphasize that the selection function is applicable to any type of binary system, e.g., star + star, or star + planet.
The selection function is defined in terms of observables (e.g., the size of the photocenter's orbit; Equation \ref{eq:photocenter}).
Although different companion luminosities and masses will result in different observables and thus different detection probabilities, \emph{Gaia} is agnostic to whether the secondary component of an unresolved binary is a star, planet, or black hole.
The purpose of the selection function is to quantify how different those detection probabilities are.

Our forward modeling approach combines analytic models and results from the individual-epoch forward modeling of \cite{Kareem} to produce fast, empirical functions to calculate the probability of a source being included in the NSS astrometric orbits catalog.

The rest of this paper is organized as follows.
In \S \ref{sec:Gaia astrometric non-single star processing and catalog} we review the processing pipeline of the \emph{Gaia} NSS astrometric orbit catalog and identify the most important steps to model.
In \S \ref{sec:Binary population synthesis} we describe the binary population synthesis model that we compare to the NSS catalog.
In \S \ref{sec:Empirical modeling of Gaia NSS model cascade} we forward model the NSS astrometric processing cascade, and the results are presented in \S \ref{sec:Results}.
Applications, strengths and shortcomings, and possible extensions of the forward model are discussed in \S \ref{sec:Discussion}.
We provide a final summary and conclusions in \S \ref{sec:Conclusions}.

% Note: I added "Summary of"
\section{\emph{Gaia} astrometric non-single star processing and catalog
\label{sec:Gaia astrometric non-single star processing and catalog}}

\begin{figure}
    \centering
    \includegraphics[width=1.0\linewidth]{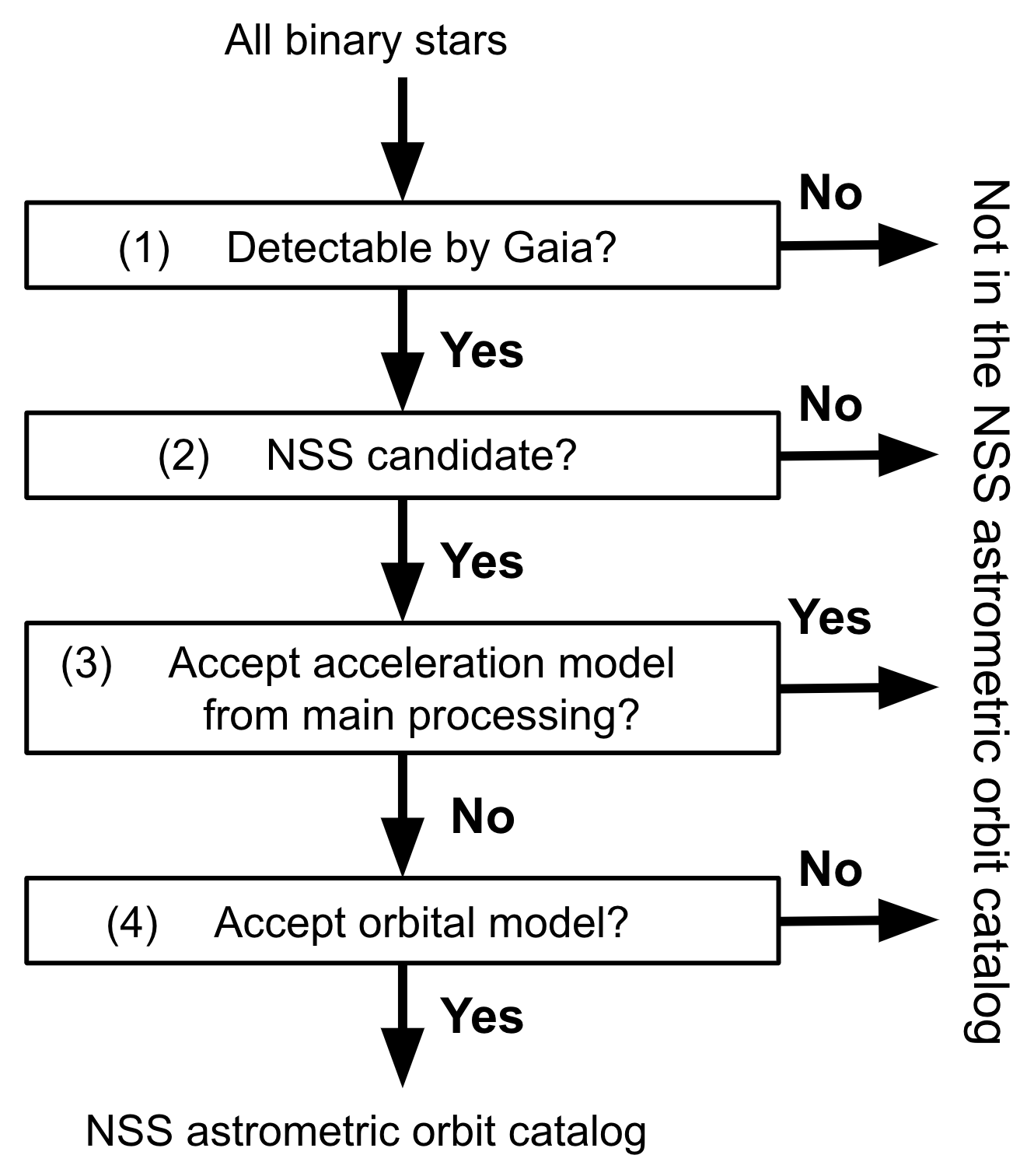}
    \caption{Summary of the NSS astrometric orbit cascade, illustrating the criteria that must be satisfied for a source to be included in the NSS astrometric orbit catalog.}
    \label{fig:cascade}
\end{figure}

The astrometric NSS processing and resultant catalog were designed with the intention of prioritizing accurate binary parameters over completeness.
This resulted in a series of many filtering steps to remove potentially spurious solutions, at the expense of many accurate solutions.
Our goal is to understand the selection effects of this filtering process, as well as determine which filters are most important to model.
The astrometric NSS processing described in \cite{Halbwachs:2023} is summarized below, and the portions most relevant to the forward modeling are highlighted in Figure \ref{fig:cascade}; see also Figure 1 of \cite{Halbwachs:2023} for a summary.

\subsection{Astrometric orbit solutions
\label{sec:Astrometric orbit solutions}}

The \emph{Gaia} DR3 NSS catalog is divided into four tables depending on the type of NSS model used; each table is further subdivided by the particular sub-model used (see \S 2.1. ``Table contents" of \cite{Arenou:2023}).
This paper is only concerned with modeling the subset of solutions that have astrometric orbits, i.e. \texttt{Orbital} and \texttt{AstroSpectroSB1} solutions from the \texttt{nss\_two\_body\_orbit} table.
We collectively refer to these as the NSS astrometric orbit solutions.
Unless otherwise stated, we will also just use the shorthand ``NSS solutions" or ``NSS catalog" to describe this dataset, since we are only concerned with astrometric models in this paper.

\subsection{Selection of stars to be processed
\label{sec:Selection of stars to be processed}}

In the astrometric NSS processing, an initial filtering was performed on the entire \emph{Gaia} DR3 \gaiasource\, catalog to define the input catalog, a subset of sources that were good NSS candidates.
Starting from $1.5 \times 10^9$ sources with single-star astrometric solutions, several filtering steps were implemented to produce the input catalog of $4.1 \times 10^6$ stars.
Only stars in this input catalog were eligible to be fit with an astrometric NSS model.
As described in \cite{Halbwachs:2023} \S 2.1, the filters are on the source's apparent magnitude, goodness of fit of a single star solution, number of visibility periods (i.e., independent observations),\footnote{Using the number of visibility periods provides a more reliable estimate for position and proper motion uncertainties than the total number of good along-scan observations \citep{Lindegren_etal:2018}.}
image parameter determination metrics, and color excess.
The filters on the image parameter determination metrics and color excess are designed to remove marginally resolved wide binaries.
\emph{Gaia} DR3 has an effective angular resolution of around 0.7 arcsec\footnote{Although \emph{Gaia} DR3 has an angular resolution of 0.18 arcsec \citep{Lindegren:2021a}, it is limited spatially by crowding, making its effective angular resolution 0.7 arcsec.}; at smaller separations the completeness of close source pairs drops \citep{GaiaEDR3:2021}.

In our forward model, we will only model the filter on the goodness of fit of a single star solution ($\ruwe > 1.4$, \S \ref{sec:Empirial ruwe model}).
The filters on apparent magnitude and visibility periods are accounted for in the \ruwe\, filter.
As \ruwe\, is a measurement of the significance of deviations from a single-star model, sources that are faint or have too few visibility periods will have astrometric uncertainties too large to have a significant deviation.
In addition, our forward model does not explicitly model the image parameter determination metrics or color excess, because the wide binaries removed by these sources will be removed by the $\ruwe > 1.4$ filter.
Using a binary population synthesis model (\S \ref{sec:Binary population synthesis}), we find this angular resolution corresponds to binaries within 2 kpc having orbital periods several centuries to millennia. 
These orbital periods are much too long to receive an astrometric orbital solution; they have $\ruwe > 1.4$ because their marginally resolved nature leads to spurious epoch astrometry.
See also \cite{Tokovinin:2023} and Appendix B of \cite{Kareem}.

\subsection{Model cascade and main processing
\label{sec:Model cascade}}

Subsequently in the astrometric NSS processing pipeline, the $4.1 \times 10^6$ stars from the input catalog were fit with a variety of astrometric NSS models.
The models were tried in the following order\footnote{We do not model the ``variability-induced movers'' branch in the processing cascade; this describes sources with astrometric motion in unresolved binaries where one source is photometrically variable.
We do not consider variable sources in our catalog so this is not relevant.}: 
\begin{itemize}
    \item variable acceleration (\texttt{Acceleration9} solution, \texttt{nss\_acceleration\_astro} table; 9 parameters),
    \item constant acceleration (\texttt{Acceleration7} solution, \texttt{nss\_acceleration\_astro} table; 7 parameters),
    \item orbital (\texttt{Orbital} solution, \texttt{nss\_two\_body\_orbit} table; 12 parameters\footnote{10 parameters were used for a circular or psuedo-circular orbit (see \S 5.2 ``Pseudo-circular and circular orbital solutions" of \cite{Halbwachs:2023}).
    15 parameters were used if the orbital solution was combined with a spectroscopic solution (\texttt{AstroSpectroSB1} solution).
    The astrometry-only and spectroscopic-only inferred parameters were only combined if they matched (see \S 5.3. ``Selection of the orbital solutions" of \cite{Halbwachs:2023}).}).
\end{itemize}
In the NSS astrometric processing pipeline, there are two steps: main processing and post-processing.
% As described in \cite{Halbwachs:2023}, these models are tried sequentially in the main processing.
If a source passes a given model's main processing criteria, it moves into the post-processing branch of that model; if a source does not pass a given model's main processing criteria, the next model in the cascade is tried.
If a source passes both a given model's main processing and post-processing criteria, it is accepted as a solution and is included in that model's catalog.
If a source does not pass the post-processing criteria, it is removed and given a single star solution.

For each model, the criteria for main processing acceptance were based on the the significance of the model and the significance of the parallax measurement.
The specific criteria are summarized in Table 1 of \cite{Halbwachs:2023}.
If an acceptable solution was found, subsequent models were not tried.
Note that the acceleration models precede the orbital model.
Thus, if an acceptable acceleration model was found, an orbital model was not tried, even if an orbital solution would have led to a better fit.

In our forward model, we model the variable acceleration (referred to as 9-parameter acceleration), constant acceleration (referred to as the 7-parameter acceleration), and the orbital components of the astrometric cascade.
For each of these three models, we model the main processing filters on the significance of the model and the parallax measurement.

Throughout this paper we discuss acceptance of acceleration and orbital models.
When we say an acceleration model is accepted or rejected, we are specifically referring to the main processing step unless otherwise stated.
This is because only the main processing branch of the acceleration models is relevant to the orbital solutions.
If an acceleration model is rejected from the main processing, it is a candidate for an orbital solution.
If an acceleration model is accepted in the main processing, it will not be a candidate for an orbital solution regardless of whether it is accepted or not in the subsequent acceleration post-processing.
% Thus, the acceleration post-processing is not necessary to understand the selection effect in the orbital solution catalog.
In contrast, when we say an orbital model is accepted or rejected, we are specifically referring to acceptance in both the main and post-processing steps.
This is because we are interested in the final astrometric orbits catalog, which requires passing the post-processing criteria.

\subsection{Orbital solution model acceptance
\label{sec:Orbital solution model acceptance}}

For sources in the NSS astrometric processing pipeline where an orbital model fit was attempted in the main processing, there are three additional conditions determining whether the solution was accepted in the post-processing and included in the final orbital solution catalog:
a criterion on the goodness-of-fit $(F_2)$, a second more stringent filter on the significance of the model, and a filter on the error of the eccentricity measurement $\sigma_e$ \citep{Halbwachs:2023}.

In our forward model, we do not model the filter on $\sigma_e$, as it removes no more than a few percent of solutions and does not bias the inferred distributions of orbital parameters.
Detailed justification is provided in Appendix \ref{app:eccentricity error}.
Visual inspection of Figure 3 of \cite{Halbwachs:2023} also suggests the filter on $\sigma_e$ has minimal impact on the final constructed catalog.

In our forward model, we also do not model the filtering on $F_2$ because we assume that good orbital solutions will have good fits. 
$F_2$ is expected to follow a normal distribution with mean 0 and standard deviation 1; even if the uncertainties are poorly characterized, $F_2 < 25$ is still a 25-$\sigma$ threshold. 
It may also allow false positives to pass through, but is unlikely to remove real solutions that have passed the other filtering steps (with the exception of binaries in hierarchical triple systems).
We provide further justification of this choice in Appendix \ref{app:Goodness-of-fit F2}.

\subsection{Final NSS astrometric orbit catalog
\label{sec:Final NSS astrometric orbit catalog}}

\begin{figure*}
    \centering
    \includegraphics[width=\linewidth]{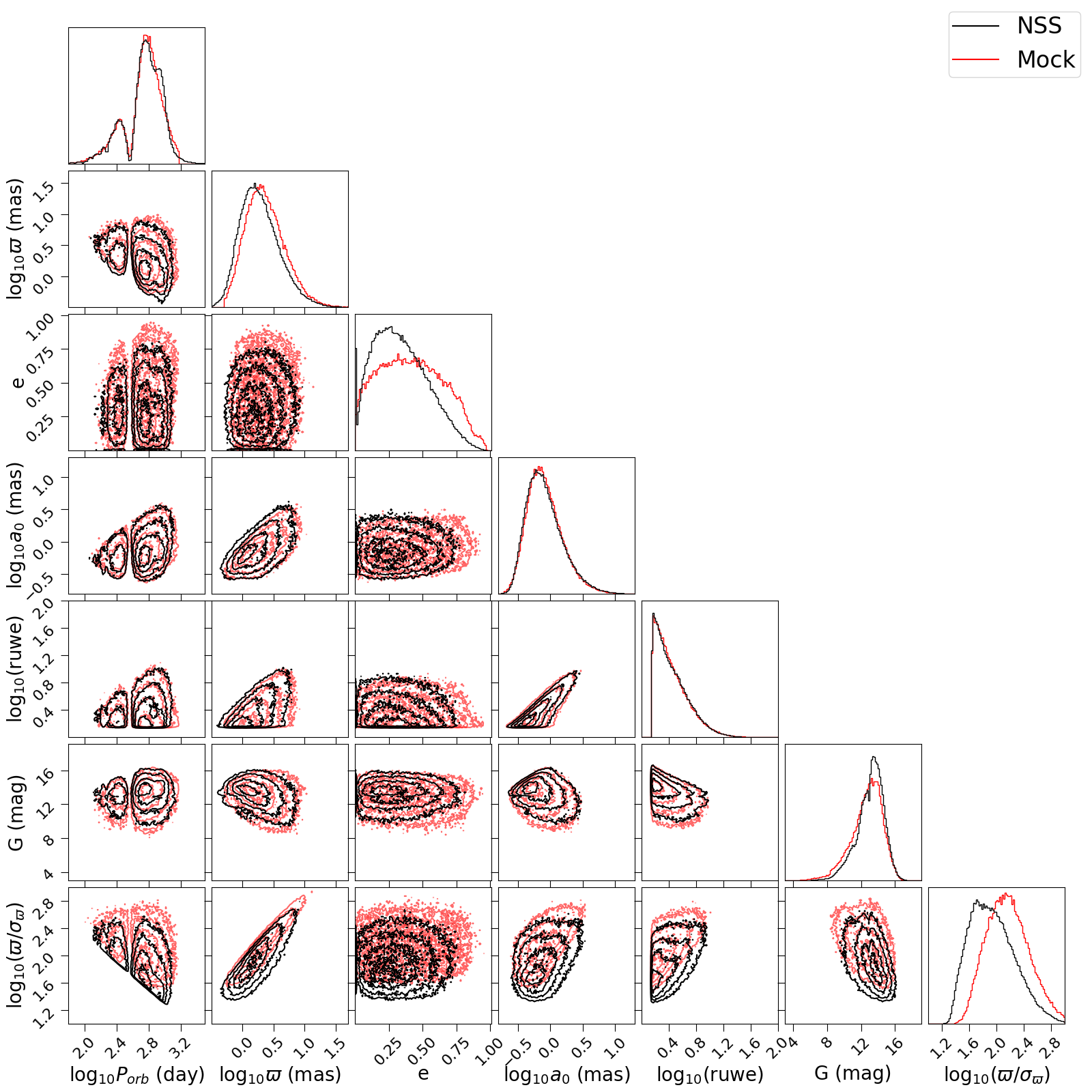}
    \caption{
    1-D and 2-D PDFs in the period $P_{orb}$, parallax $\varpi$, eccentricity $e$, photocenter orbit semimajor axis $a_0$, \ruwe, apparent magnitude $G$, and parallax over parallax error $\varpi/\sigma_\varpi$ of the actual NSS astrometric orbits catalog (black) as compared to forward model (red, described in \S \ref{sec:Binary population synthesis} -- \ref{sec:Results}).
    In general, the forward modeling does a good to reasonable job in matching the actual NSS distributions including the correlations, with the exception of the eccentricity distribution.
    \label{fig:fm_corner}
    }
\end{figure*}

The black outline in Figure \ref{fig:fm_corner} shows the distribution of orbital period, parallax, eccentricity, photocenter orbit semimajor axis, renormalized unit weight error, apparent magnitude, and parallax over parallax error for the final set of 168,065 \emph{Gaia} DR3 NSS astrometric orbit solutions.
Some of the correlations are clearly due to the filtering procedure (e.g. orbital period vs. parallax over parallax error).

\section{Binary population synthesis
\label{sec:Binary population synthesis}}

We forward model a binary population and apply the filters described in \S \ref{sec:Gaia astrometric non-single star processing and catalog} to attempt to reproduce the \emph{Gaia} NSS astrometric orbit catalog.
Note that only binary stars are modeled; single stars are not included, nor are higher-order multiple systems.

We emphasize that the population synthesis is used primarily to validate the output of the selection function, and that it is not used in determining the selection function itself.
Once the selection function has been determined, different population synthesis models can be explored to determine whether they produce different predicted numbers of detectable black holes \citep[e.g.,][]{Nagarajan:2025}, although such a study is out of the scope of the work presented here.

\subsection{Galactic model (positions, kinematics, ages, metallicities, extinction)}

We use the single-star Galactic model \texttt{galaxia} \citep{Sharma:2011} to generate stellar positions, metallicities, and ages.
\texttt{galaxia} is based on the Besan\c{c}on Galactic model of \cite{Robin:2003}, which separates the Galaxy into 4 sub-populations: the thin disk, thick disk, stellar halo, and bulge.
Each sub-population has a different stellar age, metallicity, and position distribution; see Tables 1--3 of \cite{Sharma:2011} for details. In summary, the thin disk spans ages $0 - 10$ Gyr and $\langle\textrm{[Fe/H]}\rangle$ from $-0.01$ to $-0.37$, with older components being more metal-poor; the thick disk age is 11 Gyr with $\langle\textrm{[Fe/H]}\rangle = -0.78$; the stellar halo age is 13 Gyr\footnote{We adjust the age of the stellar halo to be 13 Gyr (vs. 14 Gyr in \cite{Sharma:2011}).} with $\langle\textrm{[Fe/H]}\rangle  = -1.78$; the bulge age is 10 Gyr with $\langle\textrm{[Fe/H]}\rangle = 0$. 

We only simulate stars within 2 kpc of the Sun, as beyond that \emph{Gaia}'s sensitivity to astrometric binaries in DR3 significantly drops.
Note that within 2 kpc there are no bulge stars.
To validate this choice, we simulate all the stars in the Galaxy to determine how many solutions are potentially lost, using the forward model described later in this paper. 
Only $\approx 1\%$ of solutions are beyond 2 kpc (and all solutions are within 3.5 kpc), so limiting the simulated sample to sources within 2 kpc is justified.

\begin{figure}
    \centering
    \includegraphics[width=1.0\linewidth]{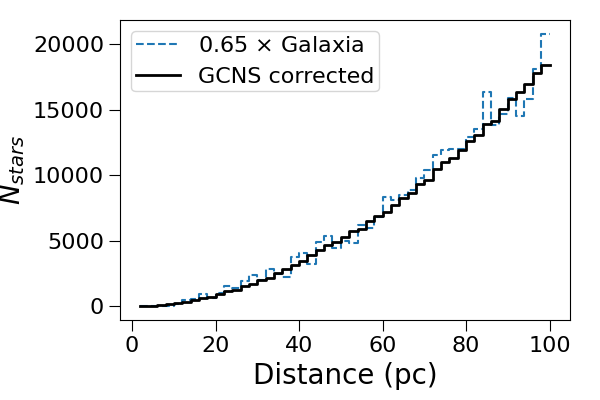}
    \caption{
    Number of sources as a function of distance from the Sun, for the Gaia Catalogue of Nearby Stars (GCNS; black line) vs. the \texttt{galaxia} simulation (blue dash).
    The GCNS is $> 92\%$ complete for stars down to $0.08 M_\odot$ within 100 pc of the Sun; to obtain the completeness corrected number of sources, we divide the number of GCNS sources in each histogram bin by 0.92.
    In order to have agreement with the completeness corrected source counts in GCNS, the number of \texttt{galaxia} sources must be reduced by a factor of 0.65 (blue line).
    \label{fig:galaxia_vs_gcns}
    }
\end{figure}

We compare the star counts predicted by \texttt{galaxia} to those in the Gaia Catalogue of Nearby Stars \citep[GCNS;][]{Smart:2021}, which is $>~92\%$ complete down to $0.08M_\odot$ within 100 pc of the Sun.
We find that \texttt{galaxia} over-predicts the number of stars within 100 pc by a factor of about 1.5 (Figure \ref{fig:galaxia_vs_gcns}).
\texttt{galaxia} predicts 0.51 million stars within 100 pc, and 1.1 billion single stars within 2 kpc.
In comparison, the GCNS contains 0.36 million sources after completeness correction within 100 pc.
Assuming this trend holds out to 2 kpc and that one GCNS source is equivalent to one \texttt{galaxia} star, we rescale the \texttt{galaxia} star counts by 0.65 to match the observed source counts in the GCNS.
After rescaling the \texttt{galaxia} star counts to match the GCNS, there are 720 million \texttt{galaxia} stars within 2 kpc.

Extinction was calculated using the 3-D dust map \texttt{mwdust} \citep{Bovy:2019}, which combines the dust maps of \cite{Marshall:2006, Drimmel:2003, Green:2019}.
\texttt{mwdust} does not have a \emph{Gaia} G-band filter, so we approximate it using the Gunn r-band filter (this is equivalent to using $A_G = 2.3 E(B-V))$.

\subsection{Stellar population model (binary properties and evolution, IMF, synthetic photometry)}

To forward model the NSS catalog, we cannot directly use  the stellar population generated by \texttt{galaxia} since it consists of only single stars.
Hence, we combine \texttt{galaxia} positions, metallicities, and ages with a binary star population based on \cite{Moe:2017}, which is a joint probability distribution in primary mass $M_1$, mass ratio $q = M_2/M_1$, orbital period, and eccentricity.
\cite{Moe:2017} infer the distribution over $0.8 < M_1 < 40 M_\odot$, $0.1 < q < 1$, $0.15 < \log P_{orb}/\textrm{day} < 8$, and $0 < e < 1$.
For reference, about 60\% of Sun-like stars are single, while this decreases to nearly 0\% for OB-type stars.

We use the Compact Object Synthesis and Monte Carlo Investigation Code \citep[COSMIC;][]{Breivik:2020} to sample this joint distribution.
We generate binary systems with primary masses spanning $0.08 < M_1 < 20 M_\odot$.
The primary masses in the distribution of \cite{Moe:2017} only go down to $0.8M_\odot$, so for the low mass primaries $0.08 < M_1 < 0.8M_\odot$, it is assumed the joint probability distribution of orbital periods and eccentricities are the same as those with primary mass $M_1 = 0.8 M_\odot$, and the binary fraction is extrapolated such that it is 0 at $M_1 = 0.08 M_\odot$, and the mass ratio distribution is truncated such that the minimum $q$ is $0.08M_\odot/M_1$.

We assume that the mass distribution of primaries and singles are drawn from a Kroupa initial mass function \citep{Kroupa:2001}.
This results in a multiplicity frequency\footnote{Following the definition of \cite{Reipurth:1993}, the number of binary systems divided by the number of binary systems and single stars.} of 0.21 and a companion probability\footnote{Again following \cite{Reipurth:1993}, the number of stars in binary systems divided by the total number of stars.} of 0.35.
This multiplicity frequency might seem significantly lower than binary frequency of $30-40\%$ typically quoted for solar-type stars \citep[e.g.,][]{Raghavan:2010, Moe:2017}.
This is because the binary statistics in \cite{Moe:2017} only extend down to $0.8 M_\odot$; to extrapolate the multiplicity statistics down to $0.08M_\odot$, COSMIC assumes the binary fraction decreases linearly from $\approx 40$\% at $0.8M_\odot$ down to 0 at $0.08 M_\odot$.
This choice of implementation is somewhat arbitrary (given that M-dwarfs appear to have a binary frequency $\sim 20\%$; e.g., \citealt{Laithwaite:2020}), but ultimately most of these sources are too faint to be included in the astrometric orbits catalog, so the model is not sensitive to the particular binary fraction assumed.
Based on an estimate using the extinction-corrected absolute magnitudes of the NSS sources, we find that main sequence M-dwarf binaries comprise $\lesssim 5\%$ of the NSS catalog.
Similarly, we do not simulate substellar companions, as they are a negligible part of the NSS catalog; \cite{Arenou:2023} find that only 1\% of NSS companions are brown dwarfs or exoplanets.

Assuming that the majority of binary systems are unresolved\footnote{Using the criterion for resolvability given by Equation (B1) in \cite{Kareem} and assuming the binary separation is equal to the semimajor axis, we find that 80\% of binary star systems will be unresolved.}, one \texttt{galaxia} star corresponds to one COSMIC stellar system.
This corresponds to 150 million binary systems within 2 kpc at the present day.
In order to properly generate the birth population, the stellar progenitors of the compact objects must be accounted for\footnote{Note that compact objects are not included in \texttt{galaxia}.}.
We use the MIST stellar evolution models \citep{Dotter:2016, Choi:2016} to evolve a population of stars to the same age distribution of the thin disk, thick disk, stellar halo, and bulge.
We then determine how many of the primary stars have or will imminently become compact objects (defined as carbon burning for massive stars, and the end of the post-AGB phase for low/intermediate mass stars), and apply a correction factor to the number of stars that need to be generated.
This correction is 7.5\% for the 5--7 Gyr old population in the thin disk, 11.5\% for the 11 Gyr old population in the thick disk, and 12.5\% for the 13 Gyr old population in the halo.
The final result is a birth population of 170 million binary systems within 2 kpc.

For binaries where the primary has reached carbon burning, we assume it has evolved into a neutron star or black hole, and the binary is disrupted by the natal kick of the neutron star or black hole.
This means there are no binary neutron stars nor black holes in the population synthesis.
This may seem strange given that one of the goals of this work is to determine the number of black holes in wide binaries in \emph{Gaia}.
However, the goal of the population synthesis is not to determine how many black hole + star binaries are produced; rather, the population synthesis is used to evaluate whether or not our framework for forward modeling the NSS astrometric catalog can reproduce the actual catalog.
Detailed population synthesis studies \citep[e.g.,][]{Yamaguchi:2018,Wiktorowicz:2020,Chawla:2022} estimate there to be $10^5 - 10^6$ black hole + star binaries in the entire Galaxy; within 2 kpc (which contains about 1\% of Galactic stars), there will be $10^3 - 10^4$ black hole + star binaries.
After applying detectability criteria, the previously mentioned studies estimate $10^1 - 10^3$ black hole + star binaries detectable by \emph{Gaia}; these are optimistic estimates as these are end-of-mission (i.e., DR4).
Thus, black hole + star binaries are a negligible part of the forward model, considering there are $O(10^5)$ astrometric solutions total.

For binaries where both the primary and secondary have reached the end of the post-AGB phase, we assume those are two white dwarfs that will not produce a \emph{Gaia}-detectable binary as both components are too faint.
For binaries where only the primary has reached the post-AGB phase, we assume these are white dwarf + star binaries.
Following the prescription of \cite{Kareem,  Yamaguchi:2024}, we assume that 10\% of these white dwarf + star binaries with initial separations between $2-6$ AU form wide post-common envelope binaries with separations that are half their initial separation.
The white dwarfs are assumed to be dark and their masses are determined from their zero-age main sequence mass using the initial-final mass relation of \cite{Kalirai:2008}.
The remaining white dwarf + star binaries are assumed to either have orbital periods too short ($\lesssim 10$ days) or too long ($\gtrsim 10^4$ days) to be detected astrometrically by \emph{Gaia} in DR3.
We also remove binaries where either star is currently or has previously overflowed its Roche lobe, under the assumption that the periods will be too short to detect astrometrically in \emph{Gaia} DR3 due to orbital shrinking.

Because of the uncertainties in the effects of interactions in binary stars and the precise properties of natal kicks in neutron stars and black holes, we do not evolve the binary systems using COSMIC; we only use COSMIC to generate the population of binary stars in their birth configuration.
The majority of binaries in the NSS astrometric catalog are main sequence + main sequence binaries that have never interacted, so treatment of evolutionary effects is less important than the modeling of the \emph{Gaia} astrometric processing cascade.

The MIST models are also used to generate synthetic \emph{Gaia} G-band photometry for the stars as a function of mass, age, and metallicity.
Again, no binary evolution or interaction is taken into account; each star is individually evolved using a single-star isochrone.

With the population simulated, we can calculate any other properties of the binaries needed.
Most important for this forward model is the angular semimajor axis of the orbit photocenter
\begin{equation}
    a_0 = a \varpi \delta_{ql},
    \label{eq:photocenter}
\end{equation}
where $a$ is the semimajor axis of the binary (in physical units), $\varpi$ is the parallax, and $\delta_{ql}$ is the photocenter scale.
The photocenter scale, the factor by which the photocenter orbit semimajor axis is smaller than the semimajor axis, is defined as
\begin{equation}
    \delta_{ql} \equiv \Bigg| \frac{q_l}{q_l+1} - \frac{l}{1+l} \Bigg|
    \label{eq:delta_ql}
\end{equation}
where $l$ is the luminosity ratio $l = L_2/L_1$ of the binary system and $q_l$ is the mass ratio $q_l = M_2/M_1$ \citep{vandeKamp:1975, Belokurov:2020}.
By definition here, the primary (component 1) is brighter, but not necessarily more massive.
The mass ratio $q$ is defined to be $0 \leq q \leq 1$.
Thus, $q_l = q$ if component 1 is both brighter and more massive, otherwise $q_l = 1/q$.
The photocenter orbit semimajor axis is maximized for dark companions (i.e., $l = 0$) and large mass ratios (i.e., $q_l \gg 1$), and is zero for equal-mass companions of the same luminosity ($l = 1$, $q_l=1$).
We note that all these luminosity-dependent quantities are specifically in the \emph{Gaia} G-band.

\subsection{\emph{Gaia}-specific observational quantities}

Finally, the population synthesis model also requires information about \emph{Gaia}'s observational properties: the number of observations and the measurement error.
To obtain an estimate for the number of visibility periods $N_{vis}$ in the population synthesis, we use the \emph{Gaia} Observation Forecast Tool (GOST)\footnote{https://gaia.esac.esa.int/gost/}, which predicts when any given part of the sky crosses \emph{Gaia}'s focal plane.
To calculate $N_{vis}$, we evaluate the number of predicted scans at 49152 evenly spaced points on the sky, discard a random 10\% of the predictions to mimic losses due to dead time\footnote{The GOST page suggests 20\% would be more appropriate, but we find 10\% gives good agreement with the \emph{Gaia} DR3 catalog}., then calculate how many of the remaining predicted observations are separated by at least 4 days.

We adopt the along-scan per-CCD centroiding error $\sigma_{AL}(G)$ from Figure 3 of \cite{Holl:2023}.
$\sigma_{AL}(G)$ is a function of the source magnitude $G$.
As described in \cite{Lindegren:2018}, for $12 \lesssim G \lesssim 17$ the formal centroiding precision is limited by photon noise (leading to Poisson errors), for $G \gtrsim 17$ background noise dominates and the error rises more quickly than Poisson, and for $G \lesssim 12$ the CCD gates limit the number of photons and the centroiding precision is independent of magnitude.
The $\sigma_{AL}(G)$ we adopt from \cite{Holl:2023} is empirically determined from \emph{Gaia} DR3 data, and is significantly larger ($> 0.3$ dex) than the formal error for bright ($G \gtrsim 12)$ sources.

\section{Empirical modeling of \emph{Gaia} NSS model cascade
\label{sec:Empirical modeling of Gaia NSS model cascade}}

With a binary population model in hand, the next step is to reproduce the modeling cascade described in \S \ref{sec:Gaia astrometric non-single star processing and catalog}.
In this section, we describe how we estimate and implement the filter that selects astrometric NSS candidates ($\ruwe > 1.4$), the effect of the acceleration models in the cascade, and the filters that select the accepted orbital solutions ($\varpi/\sigma_\varpi > 20,000 \textrm{ day}/P_{orb}$ and $a_0/\sigma_{a_0} > \max(5, 158/ \sqrt{P_{orb}/\textrm{days}})$).
Our general framework is to estimate the quantities of interest analytically, then use the simulations of \cite{Kareem} to calibrate these estimates as a function of intrinsic orbital and binary parameters (e.g., period, ecccentricity, photocenter orbit size).

\subsection{\cite{Kareem} simulations}

\cite{Kareem} generate a synthetic population of Galactic binary stars, then forward model individual 1-D astrometric measurements of these sources according to \emph{Gaia}'s scanning law and noise properties.
They then model these 1-D observations according to the NSS astrometric cascade, fitting the simulated epoch data with the various NSS astrometric models and applying the various filtering criteria.
\cite{Kareem} find good agreement between the properties of their forward modeled catalog and the actual \emph{Gaia} DR3 NSS astrometric orbits catalog.

Many of the underlying components of the population synthesis in \cite{Kareem} are similar or identical to those described in this paper; they also use \texttt{galaxia} and COSMIC to simulate the population of binary stars within 2 kpc.
Minor differences include the extinction ($A_G = 2.3 E(B-V)$ in this work vs. $A_G = 2.8 E(B-V)$ in \citealt{Kareem}), number of binaries at birth within 2 kpc (170 million in this work vs. 195 million in \citealt{Kareem}), and the white dwarf initial-final mass relation (\citealt{Kalirai:2008} in this work vs. \citealt{Weidemann:2000} in \citealt{Kareem}).
However, we emphasize that the results we derive from \cite{Kareem} in the rest of this section are independent of the details of their population synthesis. 
This is because we are considering the effects of the cascade relative to the initial catalog. 
As long as all relevant regions of parameter space are covered, then the results are independent of the specific assumptions used to generate the binary population.

\subsection{\ruwe\, model
\label{sec:Empirial ruwe model}}

The renormalized unit weight error \ruwe\, is used to quantify the goodness-of-fit of a single star model.
Specifically, it is the square root of the reduced $\chi^2_r$ (the ``uwe" in \ruwe) with a correction applied to mitigate systematics in the $\chi^2$ distribution due to color and magnitude (the ``r" in \ruwe; \citealt{Lindegren:2018}).

\subsubsection{Analytic \ruwe\, prediction
\label{sec:Analytic ruwe prediction}}

\cite{Belokurov:2020} demonstrated a close relationship between \ruwe\, and the presence of non-single stars using \emph{Gaia} DR2 data (which did not include any type of binary or NSS models).
If $\delta \theta$ is the astrometric excess noise due to fitting an unresolved binary with a single star model, then
\begin{equation}
    \delta \theta \approx \sigma_{AL} \sqrt{\ruwe^2 - 1}.
    \label{eq:delta theta ruwe}
\end{equation}

Equation \ref{eq:delta theta ruwe} assumes that the number of good observations in the along-scan direction is significantly larger than the number of degrees of freedom ($\texttt{astrometric\_n\_good\_obs\_al} \gg 5$), the binary photocenter orbit is unresolved and significant ($\ruwe > 1$), and the binarity is the dominant source of the astrometric noise (and not e.g. satellite attitude).
We confirm these are good approximations, finding a tight correlation between $\delta \theta$ and $\sigma_{AL} \sqrt{\ruwe^2 - 1}$ in the NSS astrometric catalog ($0.1 - 0.2$ dex of scatter). 

\cite{Penoyre:2020} derived an analytic relationship (their Equation 17) between the astrometric excess noise $\delta \theta$, the source magnitude $G$, photocenter orbit semimajor axis $a_0$, inclination $i$, and eccentricity $e$
\begin{equation}
    \delta \theta = a_0 \sqrt{1 - \frac{\sin^2i}{2} - \frac{3 + \sin^2i (\cos^2 \omega - 2)}{4}e^2},
    \label{eq:Penoyre20} 
\end{equation}
where $\omega$ is the argument of periapse.\footnote{In \cite{Penoyre:2020}, $\omega$ is denoted $\phi_\nu$ and is called the azimuthal viewing angle of the binary. 
We have chosen to use the more standard notation of the Campbell elements.}
Note that the derivation of Equation \ref{eq:Penoyre20} assumes that there are many observations sampling the full orbital period of the binary during the survey, such that the observations can be time-averaged over the orbit.

\begin{figure*}
    \centering
    \includegraphics[width=\linewidth]{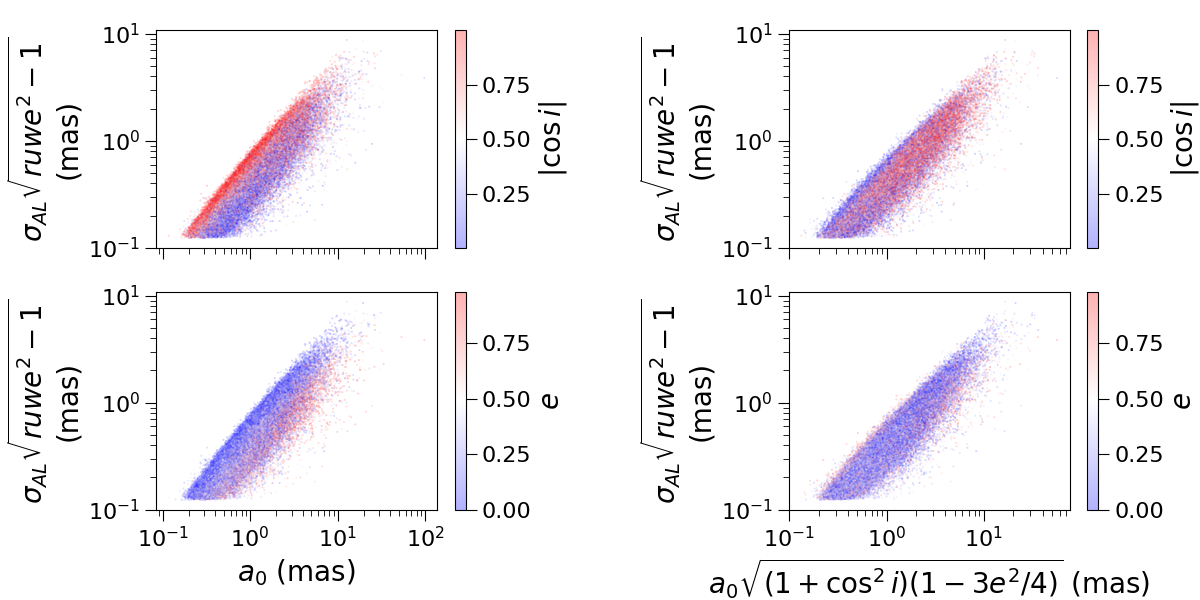}
    \caption{Using the NSS astrometric orbits catalog, we validate the scaling relations between the approximation of the astrometric excess noise $\delta \theta \approx \sigma_{AL} \sqrt{\ruwe^2 - 1}$ and the semimajor axis of the photocenter orbit $a_0$,  eccentricity $e$, and orbital inclination $i$.
    The left column shows the approximation of $\delta \theta$ vs. $a_0$, with the color of the point denoting $|\cos i|$ (top panel) or eccentricity of the orbit (bottom panel).
    There is a linear trend between $a_0$ and $\delta \theta$ where the slope is less than unity, as expected.
    In addition, there is a clear gradient in the colors of the points: averaging over the orbit, face-on and circular orbits $(|\cos i| \approx 1, e \approx 0)$ have larger astrometric excess noise at fixed $a_0$ than edge-on and highly eccentric orbits $(|\cos i| \approx 0, e \approx 1)$.
    The right column is the same as the left column, but includes the dependence of $e$ and $i$ on $\delta \theta$, i.e. shows the approximation of $\delta \theta$ vs. $a_0 \sqrt{(1 + \cos^2 i) (1 - 3e^2/4)}$. 
    The trends in the colors of the points mostly disappear (the $|\cos i|$ and $e$ dependence), validating the scaling relation.
    There are still some minor trends visible in $|\cos i|$, but \cite{Arenou:2023} showed that the sensitivity of circular orbits deviates slightly from the expected $1 + \cos^2 i$, so this is not surprising. 
    Note that there are very few red points in the bottom panels because there are very few high-eccentricity orbits. 
    \label{fig:validate_ruwe_a0_e_i_scaling}
    }
\end{figure*}

For an isotropic distribution of $\omega$, the average value of $\cos^2$$\omega$ is 0.5.
Fixing $\cos^2$ $\omega$ to 0.5, the astrometric excess noise can be re-written
\begin{equation}
    \delta \theta = a_0 \sqrt{\frac{1}{2} \left(1 + \cos^2 i\right)\left(1 - \frac{3e^2}{4}\right)}.
    \label{eq:delta theta a0}
\end{equation}
Combining Equations \ref{eq:delta theta ruwe} and \ref{eq:delta theta a0} results in a formula for \ruwe\, as a function of source magnitude, photocenter orbit semimajor axis, eccentricity, and inclination:
\begin{equation}
    \ruwe = \sqrt{1 + \frac{1}{2} \left( \frac{a_0}{\sigma_{AL}(G)} \right)^2 \left(1 + \cos^2 i \right) \left(1 - \frac{3}{4}e^2  \right)}.
    \label{eq:ruwe_G_a0_e_i}
\end{equation}
Inspection of Equation \ref{eq:ruwe_G_a0_e_i} shows that face-on orbits ($|\cos i| \approx 1$) have larger \ruwe\, than edge-on orbits ($|\cos i| \approx 0$), and circular orbits ($e \approx 0$) have larger \ruwe\, than highly eccentric orbits ($e \approx 1$).
If the orbit is perfectly face on and circular ($i=0$, $e=0$), this suggests that the astrometric excess noise is equivalent to the semimajor axis of the photocenter orbit (note that $\delta \theta \leq a_0$).
Thus, on average, filtering on a minimum value of \ruwe\, favors circular and face-on orbits.
However, \ruwe\, depends more strongly on inclination than eccentricity: the term containing inclination $1 + \cos^2 i$ spans the interval [1, 2], but the term containing eccentricity $1 - 3e^2/4$ only spans the interval [0.25, 1].
We also visually validate the scaling between the astrometric excess noise and photocenter orbit semimajor axis, inclination, and eccentricity in Figure \ref{fig:validate_ruwe_a0_e_i_scaling}.

\subsubsection{Empirical correction to analytic \ruwe\, prediction
\label{sec:Empirical correction to analytic ruwe prediction}}

The predicted astrometric excess noise in Equation \ref{eq:delta theta a0} and \ruwe\, in Equation \ref{eq:ruwe_G_a0_e_i} will generally be overestimates of the astrometric excess noise and \ruwe\, reported by \emph{Gaia}.
The derivation of Equation \ref{eq:delta theta a0} assumes that the binary signal has not been erroneously attributed to other astrometric terms (i.e., position, parallax, proper motion, or satellite pointing noise).
\cite{Penoyre:2020} predicted that this erroneous attribution would happen frequently, often lowering the reported astrometric excess noise and \ruwe.
With the NSS catalog, we confirm that the predicted astrometric excess noise $\delta \theta$ tends to be larger than the actual $\delta \theta$ in the catalog by $\approx 60 - 70\%$. % (Figure \ref{fig:aen_vs_predict}).

We calculate an empirical correction to rescale the analytic \ruwe\, prediction to match the \ruwe\, measured by fitting a single-star model to the simulated observations in the catalog of \cite{Kareem}.
We use a catalog that includes a random subset of the birth population of binaries for which the orbital parameters are known and $\ruwe$ is calculated, but \emph{before} the $\ruwe > 1.4$ filter has been applied.
For each simulated source in this catalog, we calculate $\delta \theta_{analytic}$ (the expected astrometric excess noise using Equation \ref{eq:delta theta a0}) and $\delta \theta_{fit}$ (the astrometric excess noise inferred from the \ruwe\, measured in the fitting procedure; Equation \ref{eq:delta theta ruwe}).
The ratio $\delta \theta_{fit}/\delta \theta_{analytic}$ is the rescaling factor required to correct the analytic prediction of $\delta \theta$ to account for the erroneous attribution that occurs in the fitting process.
Within each bin, we only consider sources with $\sigma_{AL} < a_0$ (as otherwise their orbits are not detectable), and sources with $\ruwe > 1.01$ (as otherwise the predicted value of $\delta \theta$ will be imaginary).

\begin{figure}
    \centering
    \includegraphics[width=\linewidth]{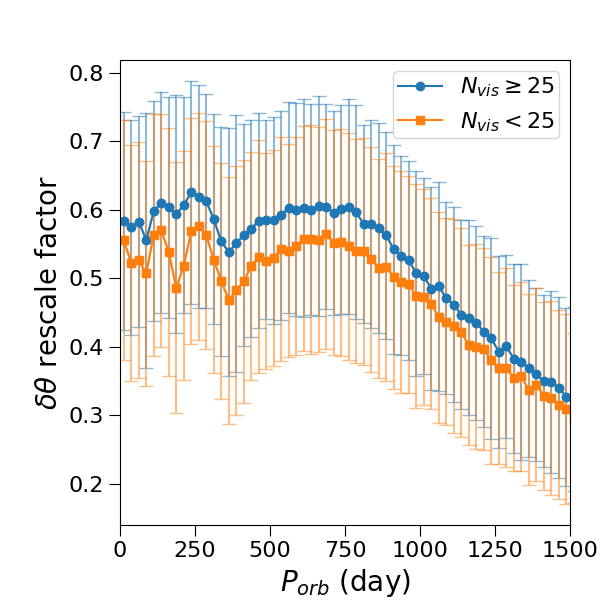}
    \caption{Comparison of the astrometric excess noise $\delta \theta$ for sources in the synthetic catalog of \cite{Kareem} as inferred by their \ruwe, compared to the $\delta \theta$ as predicted by Equation \ref{eq:delta theta a0}.
    At longer orbital periods, the rescale factor gets smaller; when a orbit is only partially observed, it can be more easily be fit a single-star model, decreasing the astrometric excess noise.
    In addition, there are clear dips at orbital periods that are harmonics of 1 year or related to the scanning law.
    Finally, when the number of visibility periods is higher, the rescale factor is slightly larger (closer to unity) and the dips at the harmonics of 1 year are smaller.
    This is because when the orbit is better sampled, the solution is less prone to degeneracies that lead to incorrect solutions.
    \label{fig:aen_vs_predict}
    }
\end{figure}

In order to account for the dependence of this correction factor on the orbital period and number of observations, we split the simulated catalog solutions from \cite{Kareem} into orbital period bins of 25 days, spanning 0 to 1500 days.
We further subdivide each period bin based on the number of visibility periods used: $N_{vis} < 25$, and $N_{vis} \geq 25$.
These bins were chosen since $N_{vis}$ is a bimodal distribution for the NSS astrometric orbits catalog, with peaks at $N_{vis} = 18$ and 27.
For each $N_{vis}$ bin, a Gaussian is fit to the ratio $\delta \theta_{fit}/\delta \theta_{analytic}$.
The mean of this Gaussian is taken to be the correction factor for that range of $P_{orb}$ and $N_{vis}$ (Figure \ref{fig:aen_vs_predict}).
To determine the correction for a particular orbit, we interpolate across the $P_{orb}-N_{vis}$ grid to find the correction factor to $\delta \theta$, allowing the accurate prediction of \ruwe\, without actually having to generate and fit mock observations.

We treat viewer-dependent quantities (e.g., argument of periastron, inclination, time of passage through periastron) as nuisance parameters; our probabilistic function is effectively integrating over these quantities.
Although there are selection effects due to viewing angle (e.g., face-on orbits are more likely to show a larger astrometric signal than an edge-on orbit), these are not intrinsic binary properties.
We are not concerned with characterizing these detection biases since these are randomized across all binaries. 
This enables us to understand the trends analytically, and the qualitative and quantitative effects of the fitting and modeling procedures.

As shown in Figure \ref{fig:aen_vs_predict}, this correction is a strong function of orbital period.
For $P_{orb} \gtrsim 1000$ days, the key assumption used to derive Equation \ref{eq:Penoyre20} (having a fully-observed orbit) is violated; even for $P_{orb} \approx 1000$ days, this assumption may not hold if the orbit is not well-sampled for any reason (e.g., a highly eccentric orbit, or low number of visibility periods).
At very long orbital periods $P_{orb} \gg 1000$ days, such a small fraction of the orbit will be covered by the Gaia DR3 temporal baseline that Equation \ref{eq:Penoyre20} is no longer a valid approximation.
The number of visibility periods used in the solution ($N_{vis}$) also affects the size of the erroneous attribution.
For larger $N_{vis}$, the average size of the erroneous attribution is slightly smaller because more independent measurements better constrain the true orbit.

We emphasize that we only model orbital periods shorter than 1500 days.
This is approximately the range over which there is a still a useful correlation between the prediction of Equation \ref{eq:Penoyre20} and the true \ruwe.
This choice is justified because in the NSS catalog, only 1\% of solutions have $P_{orb} > 1500$ days; this is a negligible fraction of solutions compared to the other uncertainties in the forward modeling.
We recommend using the methodology of \cite{Kareem} and working through the full framework of epoch modeling when trying to understand $P_{orb} > 1500$ day solutions.

\subsection{Removal of acceleration solutions candidates
\label{sec:Removal of acceleration solutions candidates}}

Although understanding the selection effects of the acceleration solutions catalog is not the focus of this work, it must be modeled because it precedes the orbital solution modeling in the cascade.
This is because a source that has an acceleration solution accepted in the main processing is removed from being a candidate for an orbital solution (Figure \ref{fig:cascade}).
As described in \S \ref{sec:Model cascade}, acceleration solutions are tried before orbital solutions in the model cascade, and if an acceptable acceleration solution is found, an orbital solution is not searched for, even if an orbital solution would have resulted in a better fit.
Thus, long-period binaries are more likely to be satisfactorily fit with an acceleration model than binaries with orbital periods short enough to be covered multiple times over DR3, and are likely to not be included in the orbit catalog.

The intrinsic binary parameters that most strongly determine whether an acceleration solution is accepted are the orbital period $P_{orb}$ and the size of the photocenter orbit semimajor axis normalized by the measurement error $a_0/\sigma_{AL}$\footnote{Although a source's positional uncertainty is determined by both its brightness and number of visibility periods, it is primarily a function of the brightness; thus, at the moment, we do not model the number of visibility periods.}. Eccentricity does not significantly affect the acceptance probability; see Appendix \ref{app:Acceleration modeling and orbital properties} for details.

\begin{figure}
    \centering
    \includegraphics[width=\linewidth]{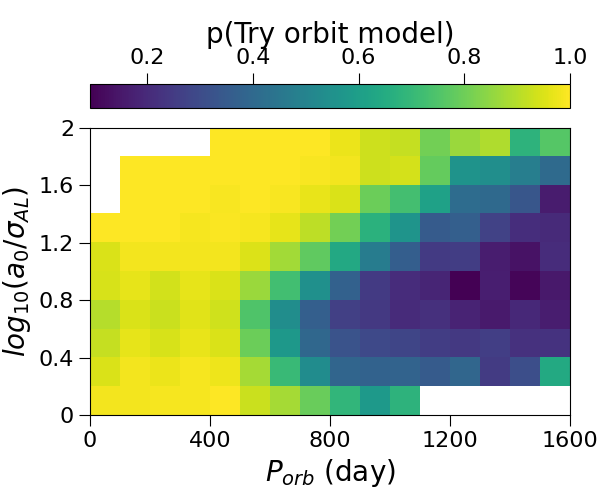}
    \caption{Probability than an orbit model is tried for an NSS candidate, as a function of orbital period $P_{orb}$ and the photocenter orbit size $a_0$, normalized by the scan error $\sigma_{AL}$.
    Nearly all sources with orbital periods $P_{orb} \lesssim 500$ days are candidates for orbital solutions.
    However, for $P_{orb} \gtrsim 500$ days, the longer orbit solutions are increasingly unlikely to be candidates for an orbital solution, especially those with less significant photocenter orbit sizes.
    This is because these solutions are lost to the acceleration portion of the model cascade.}
    \label{fig:p_try_orb}
\end{figure}

We use the simulated catalog of \cite{Kareem} to construct an empirical probabilistic function to determine which sources will be candidates for orbital solutions. 
We start with all the NSS sources that pass the initial filtering step $(\ruwe > 1.4)$ and bin them by $P_{orb}$ and $a_0/\sigma_{AL}$. 
Within each $P_{orb}-a_0/\sigma_{AL}$ bin, we calculate the fraction of sources which are candidates for orbital solutions (sources where neither a 9-parameter or 7-parameter acceleration solution is accepted in the main processing).
This is shown in Figure \ref{fig:p_try_orb}.
To find the probability that a source with $\ruwe > 1.4$ is a candidate for an orbital solution, we interpolate across this  $P_{orb}-a_0/\sigma_{AL}$ grid.

The shapes of the high and low probability regions in Figure \ref{fig:p_try_orb} can be understood as follows.
A low probability of trying an orbital model is equivalent to a high probability of accepting an acceleration solution, and vice versa.
Acceleration solutions are unlikely to be accepted at short orbital periods $P_{orb} \lesssim 500$ days because \emph{Gaia} DR3 will have covered at least two full orbits, and an acceleration model will generally be a very poor fit; thus, the probability of trying an orbital solution is very high for $P_{orb} \lesssim 500$ days.

At longer orbital periods $P_{orb} \gtrsim 500$ days, the probability of accepting an acceleration solution is maximal around the ``sweet spot" of $\log_{10}(a_0/\sigma_{AL}) \approx 0.8$ and $P_{orb} \approx 1300$ days.
If $\log_{10}(a_0/\sigma_{AL}) \lesssim 0.8$, then the detected acceleration will be noisy and less significant, making it less likely to pass the acceleration acceptance criterion on model significance.
On the other hand, if $\log_{10}(a_0/\sigma_{AL}) \gtrsim 0.8$, the astrometric signal will be much stronger, but the suitability of the acceleration model will depend on the orbital period. 
For fractionally covered orbits $P_{orb} \gtrsim 1000$ days, an acceleration model will be a good fit.
However, for orbits $P_{orb} \approx 500 - 1000$ days where a full orbit or more has been covered, measurement noise is needed to mask the model mismatch.
This explains the diagonal boundary separating the high and low probability regions for $P_{orb} \gtrsim 500$ days and $\log_{10}(a_0/\sigma_{AL}) \gtrsim 0.8$ in Figure \ref{fig:p_try_orb}.

As an independent test of whether or not longer period orbital solutions are preferentially assigned acceleration solutions, we cross-match the \nssacc\, catalog to external binary catalogs to empirically estimate how many orbital solutions are lost to acceleration solutions in the model cascade (Appendix \ref{app:cross-match to sb9}).
After accounting for the number of solutions at a given orbital period, we confirm that the fraction of acceleration solutions increases at longer periods, and the fraction of orbital solutions decreases at longer periods.

\subsection{Orbital model acceptance criteria: parallax and photocenter semimajor axis error \label{sec:Empirical parallax error}}

\begin{figure*}
    \centering
    \includegraphics[width=1.0\linewidth]{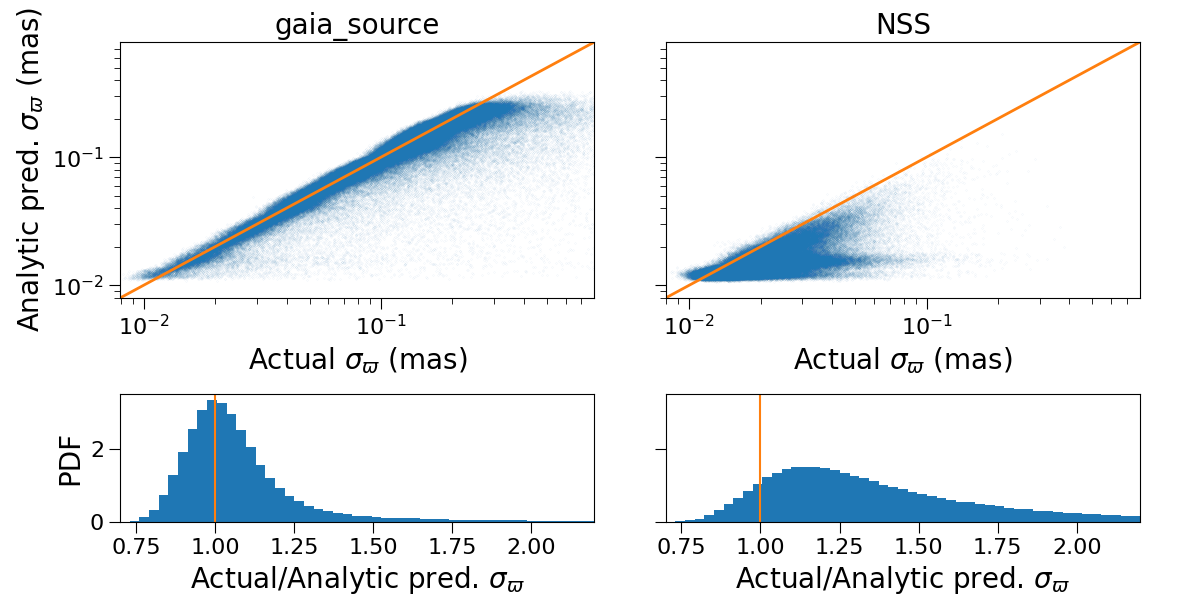}
    \caption{Parallax errors in the \texttt{gaia\_source} (left) and NSS (right) catalogs.
    The analytic prediction (Equation \ref{eq:sigma}) for the parallax errors and the true catalog errors are tightly correlated in the case of \texttt{gaia\_source} (top panel); there is a tail of underestimated errors that we attribute to non-single sources being erroneously fit with single-star models (bottom panel).
    On the other hand, the analytic prediction and the true catalog errors are highly discrepant in the \texttt{NSS} catalog; the correlation is weak (top panel) and the analytic estimates systematically underestimate the errors (bottom panel).
    This is due to erroneous attribution of errors due to correlations between the parallax and orbit model parameters such as orbital period.
    This poor agreement necessitates an empirical correction to $\sigma_\varpi$ and $\sigma_{a_0}$ in order to provide accurate uncertainties for NSS sources (Figure \ref{fig:orb_err_rescale}).
    \label{fig:parallax_error}}
\end{figure*}

In order to model the selection criteria of the orbital solutions, we need an estimate of the orbit significance $s_{orb} = a_0/\sigma_{a_0}$ and the parallax over error $\varpi/\sigma_\varpi$.
The parameters $a_0$ and $\varpi$ are directly calculated from the binary source's properties; we thus need a way to estimate their uncertainties.

\subsubsection{Analytic $\sigma_\varpi, \sigma_{a_0}$ prediction \label{sec:Analytic parallax error prediction}}

The parallax error $\sigma_\varpi$ 
and photocenter orbit semimajor axis error $\sigma_{a_0}$ are expected to be proportional to the individual measurement errors, and decrease with the square root of the number of observations: $\sigma_{AL}(G)/\sqrt{N_{vis}}$.
Using a random subset of the single-star fits in the $\texttt{gaia\_source}$ catalog, we find that the constant of proportionality is 0.48 for the parallax errors (Figure \ref{fig:parallax_error}).
Because we cannot measure the constant of proportionality for $\sigma_{a_0}$ using the $\texttt{gaia\_source}$ catalog since it consists of single-star fits, we indirectly infer it by inspecting the distribution of $\log_{10}(\sigma_\varpi/\sigma_{a_0})$ in the NSS catalog.
It is symmetric with its peak at $\sigma_\varpi/\sigma_{a_0} \approx 1$, so we assume that the constant of proportionality for $\sigma_{a_0}$ is the same as for $\sigma_\varpi$.
We thus adopt for our analytic error model
\begin{equation}
    \sigma_\varpi, \sigma_{a_0} = \frac{0.48 \sigma_{AL}(G)}{\sqrt{N_{vis}}}.
    \label{eq:sigma}
\end{equation}
Having the constant of proportionality exactly correct is not crucial, since we will be empirically rescaling it, as discussed in the next section.

\subsubsection{Empirical correction to analytic $\sigma_\varpi, \sigma_{a_0}$ prediction
\label{sec:Empirical correction to analytic error prediction}}

In practice, Equation \ref{eq:sigma} will underestimate the true $\sigma_\varpi$ and $\sigma_{a_0}$ for sources in the NSS astrometric orbits catalog (Figure \ref{fig:parallax_error}).
Analogous to the discussion in \S \ref{sec:Empirical correction to analytic ruwe prediction}, this underestimation is due to the erroneous correlations between the parallax, position, proper motion and orbital parameters when performing the model fit.
This erroneous error cannot be derived analytically, as it depends on the modeling of individual epoch observations.
Thus, we construct an empirical $\sigma_\varpi$ and $\sigma_{a_0}$ correction using the simulations of \cite{Kareem}, analogous to the empirical $\ruwe$ correction in \S \ref{sec:Empirical correction to analytic ruwe prediction}.

We analyze $3.3 \times 10^5$ sources from the simulated cascade of \cite{Kareem} that are fit with an orbital solution and that have reliably measured periods and eccentricities (i.e., the inferred $P_{orb}$ and $e$ are consistent within $3\sigma$ of the true $P_{orb}$ and $e$).
Although the simulated cascade of \cite{Kareem} fits $2.2 \times 10^6$ orbital solution candidates, we only model the 15\% of these sources with reliable fits as the remainder are spurious solutions that are not going to have accurate error estimates. 
Because \cite{Kareem} are modeling the full \emph{Gaia} DR3 cascade, they generate spurious solutions which are removed by the $F_2$ cut.
These spurious solutions are not present in our analytic model, as they arise purely from the epoch modeling.

We bin the reliably fit sources by their orbital periods and eccentricities.
Within each bin, we fit a Gaussian to the distribution of $\log_{10}(\sigma_{\varpi,fit}/\sigma_{\varpi,analytic})$, where $\sigma_{\varpi,fit}$ is the parallax error as measured when modeling epoch astrometry in the catalog of \cite{Kareem}, and $\sigma_{\varpi,analytic}$ is the parallax error predicted from Equation \ref{eq:sigma}.
We then interpolate over this grid of values to sample the means and standard deviations, which are then sampled to assign a correction to $\sigma_{\varpi,analytic}$.
We perform the same procedure on $\sigma_{a_0,fit}$ and $\sigma_{a_0,analytic}$ to determine a correction for the photocenter orbit semimajor axis $\sigma_{a_0,analytic}$.

\begin{figure*}
    \centering
    \includegraphics[width=\linewidth]{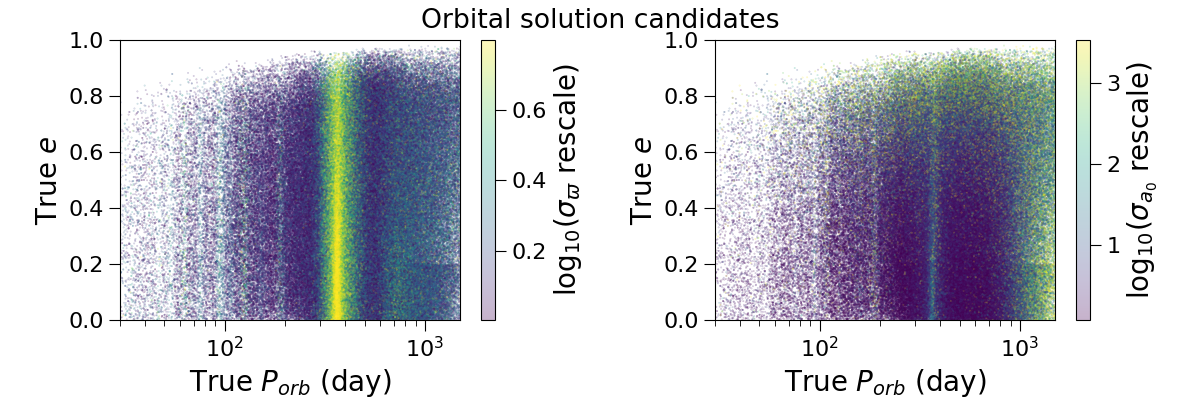}
    \caption{Difference in predicted parallax error $\sigma_\varpi$ (left) and photocenter orbit semimajor axis $\sigma_{a_0}$ (right), as a function of orbital period, eccentricity, and photocenter orbit semimajor axis for a subset of bright, well-measured sources.
    The required rescaling of $\sigma_\varpi$ is primarily driven by orbital period, with it being significantly under-predicted at $P_{orb} \approx 1$ year, as well as being somewhat under-predicted at longer orbital periods ($P_{orb} \gtrsim 550$ days).
    The rescaling of $\sigma_{a_0}$ is more complex, with dependence on both orbital period and eccentricity.
    In addition to being under-predicted at $P_{orb} \approx 1$ year and $P_{orb} \gtrsim 1000$ day, it is also significantly underpredicted at high eccentricities ($e \gtrsim 0.6$).
    \label{fig:orb_err_rescale}}
\end{figure*}

The empirical correction is shown in Figure \ref{fig:orb_err_rescale}.
The correction for $\sigma_\varpi$ is primarily a function of the orbital period.
Equation \ref{eq:sigma} underpredicts $\sigma_\varpi$ by a factor of $10^{0.7} \approx 5$ for $P_{orb} \approx 1$ year, due to the degeneracy between $P_{orb} = 1$ year orbits and parallax.
The correction for $\sigma_{a_0}$ is both a function of eccentricity and orbital period.
Solutions with either long orbital periods $P_{orb} \gtrsim 1000$ day or high eccentricities $e \gtrsim 0.6$ have $\sigma_{a_0}$ corrections around two orders of magnitude larger than solutions with lower eccentricity and shorter orbital period, excluding orbits with $P_{orb} \approx 1$ year.
This demonstrates that the significance of the orbital solution is a function of the properties of the orbit itself.

We emphasize that these empirical corrections to $\sigma_{a_0}$ and $\sigma_\varpi$ are only for solutions that have not received an acceleration solution in the NSS cascade, as they are derived only using sources that are fit with an orbital solution.
We also emphasize that the corrections are specific to the observing strategy of \emph{Gaia} DR3, and that they cannot be generally applied to observations that have a different temporal sampling, including \emph{Gaia} DR4.

\section{Results 
\label{sec:Results}}

\subsection{Mock NSS astrometric orbits catalog}

We investigate the effects of the filtering steps on the number of solutions and the distributions of orbital and binary properties.
We also present the resulting parameter distribution of the forward model.

\subsubsection{Number of solutions}

Table \ref{tab:filter number} summarizes the effect of the filters on the number of orbital solutions.
The forward model begins with 150 million binaries within 2 kpc of the Sun at the present day.
Even before applying the filters discussed in the previous section, many binaries are simply not detectable because they are too faint ($G \gtrsim 21$) to be detected by \emph{Gaia}, or their orbital periods are very long ($P_{orb} \gtrsim$ 30 years).
In order to generate a more realistic estimate on the number of potentially detectable \emph{Gaia} binaries, we calculate the number of binaries with $P_{orb} < 1500$ days and $G < 19$ mag, which represents the range of orbital periods and magnitudes spanned in the NSS astrometric orbits catalog; 12 million binaries satisfy these criteria.
Ultimately, after applying all the filtering cuts, we find $2.6 \times 10^5$ binaries will be included in the NSS catalog.

This is about 50\% more than the number of sources actually in the NSS astrometric orbits catalog (168,065).
There are several possible explanations for this overestimate.
For example, the initial binary population produced by the binary population synthesis was incorrect in some way, e.g., the rescaling factor in \S \ref{sec:Binary population synthesis} based on the GCNS was overestimated due to an overdensity of stars in the solar neighborhood, the orbital period distribution of the binaries was incorrect and produced too many systems with $P_{orb} \approx 100 - 1000$ days, the extinction was underestimated, or some binaries were disrupted by stellar encounters or cluster dispersal.
Nonetheless, the forward model is able to reproduce the total number of sources within a factor of two, which will still provide useful constraints in population studies.

\begin{deluxetable*}{p{0.35 \linewidth} | p{0.15\linewidth} | p{0.5 \linewidth}}
% \begin{deluxetable*}{l | l | l}
\tablecaption{Selection effect of NSS cascade filters}
\label{tab:filter number}
\tablehead{
    \colhead{Filter} & 
    \colhead{No. of binaries (\%)} &
    \colhead{Selection effect}
    }
\startdata
    None (all binaries) & $1.5~\times~10^8$~(100\%) & N/A \\
    \hline
    NSS potentially detectable ($P_{orb} < 1500$ day, $G < 19$ mag) & $1.2~\times~10^7$~(8\%) & Preferential selection of primaries with $M_1 \approx 1 M_\odot$ \\
    \hline
    NSS catalog candidates: \ruwe\, $> 1.4$ & $1.3 \times 10^6$ (0.9\%) & Preferentially selects $P_{orb} \approx 10^{3 \pm 1}$ days \\
    \hline
    Orbital solution candidates (i.e., not selected to acceleration solution candidates): $s_{7,9} < 12$ or $\omega/\sigma_\omega > 2.1 s_{7,9}^{1.05}$ & $7.4 \times 10^5$ (0.5\%) & Preferentially removes $P_{orb} \gtrsim 500$ days, increasing from 0\% at $P_{orb} \approx 500$ days up to 85\% at $P_{orb} \approx 2000$ days \\ %; 2) moderate eccentricity solutions $e \approx 0.2 - 0.8$} \\
    \hline
    Orbital solutions ($\varpi/\sigma_\varpi > 20,000 \textrm{ day}/P_{orb}$, $a_0/\sigma_{a_0} > \max(5,158/\sqrt{P_{orb}/\textrm{day}}$) & $2.6 \times 10^5$ (0.2\%) & Preferentially removes 1) high eccentricity $e \gtrsim 0.6$ solutions; 2) short-period $P_{orb} \lesssim 500$ days; 3) $P_{orb} \approx 1$ year solutions \\
\enddata
\tablecomments{
Steps in the NSS astrometric processing cascade, the resulting number of binaries, and a summary of the selection effect introduced by the filter.
Note that ``NSS potentially detectable" is not a filter described in \cite{Halbwachs:2023}.
It is included as a proxy for the fraction of the total binary population that could realistically receive an astrometric solution; most binaries are too faint to be detected by \emph{Gaia} or have orbital periods significantly longer than the $\approx 3$ year time baseline of DR3.}
\end{deluxetable*}

\subsubsection{Distribution of binary and orbital parameters 
\label{sec:Distribution of binary and orbital parameters}}

\begin{figure*}
    \centering
    \includegraphics[width=\linewidth]{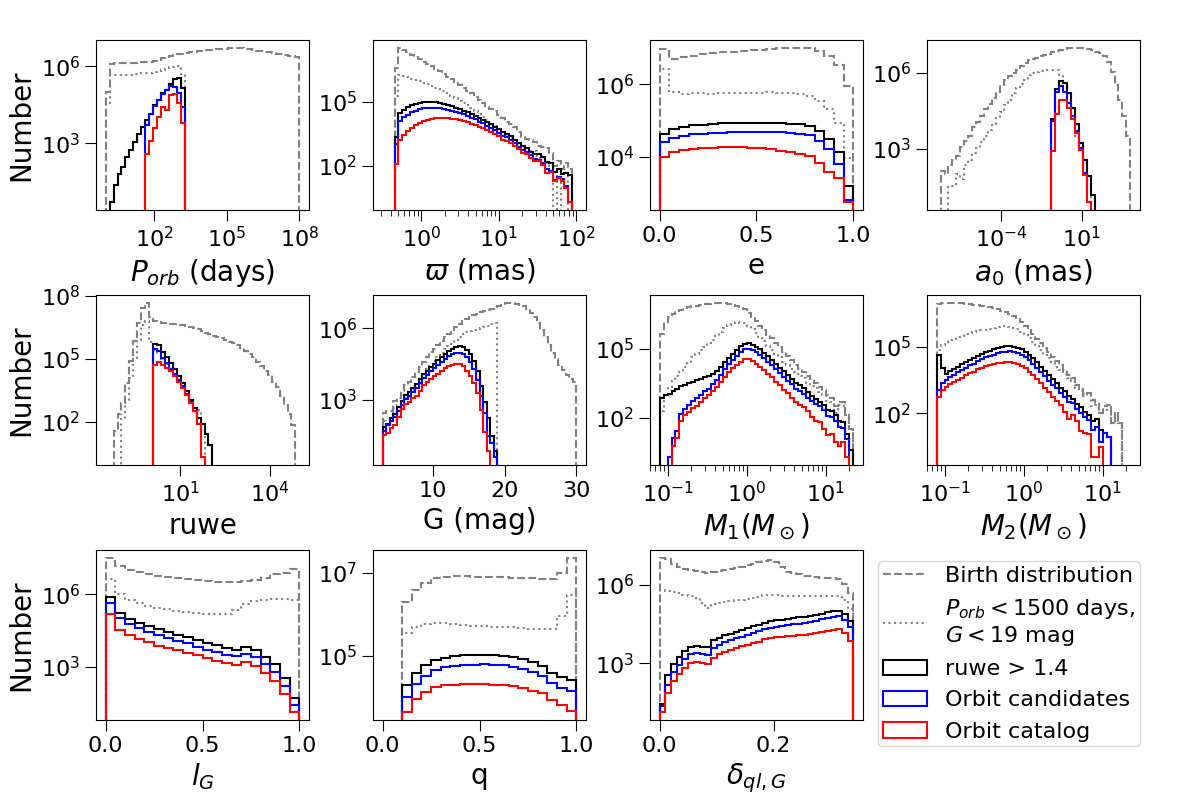}
    \caption{
    Effect of filtering on the distribution of orbital period $P_{orb}$, parallax $\varpi$, eccentricity $e$, \ruwe, magnitude $G$, primary and secondary mass $M_1, M_2$, luminosity ratio $l_G$, mass ratio $q$, and photocenter scale $\delta_{ql}$.
    The dashed gray line shows the birth distribution from \cite{Moe:2017}.
    The dotted gray line shows the distribution of systems that have orbital periods $P_{orb} < 1500$ days and $G < 19$ mag, which is the range spanned by the NSS data and a rough proxy for systems that could be characterized by \emph{Gaia}.
    The magnitude cut preferentially removes distant and low-mass binaries. 
    It causes the primary mass distribution to peak roughly around $0.8 M_\odot$ (when binned logarithmically).
    The solid black line shows sources that are NSS model candidates ($\ruwe > 1.4$), the blue line shows sources that are orbital model candidates (i.e., NSS model candidates that are not candidates for an acceleration solution), and the red line shows the sources that are in the final astrometric orbits catalog.
    Zoom-ins of the $P_{orb}$ and $e$ panels are shown in Figure \ref{fig:fm_cuts_period_ecc}.
    \label{fig:alive_bright_cuts_2}
    }
\end{figure*}

\begin{figure*}
    \centering
    \includegraphics[width=0.49\linewidth]{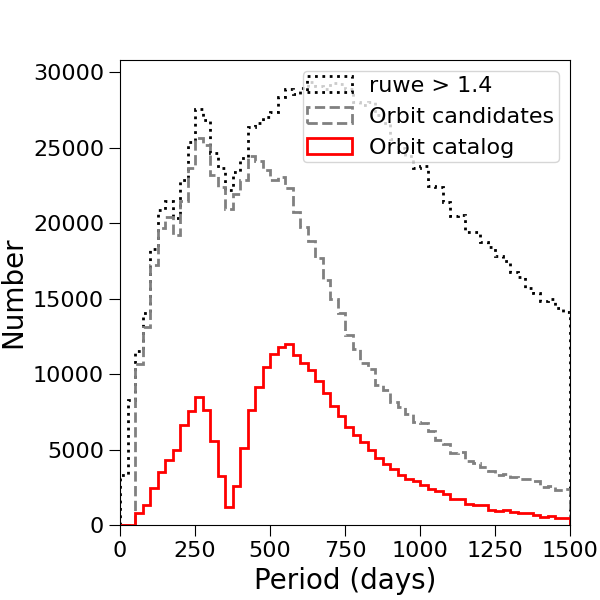}
    \includegraphics[width=0.49\linewidth]{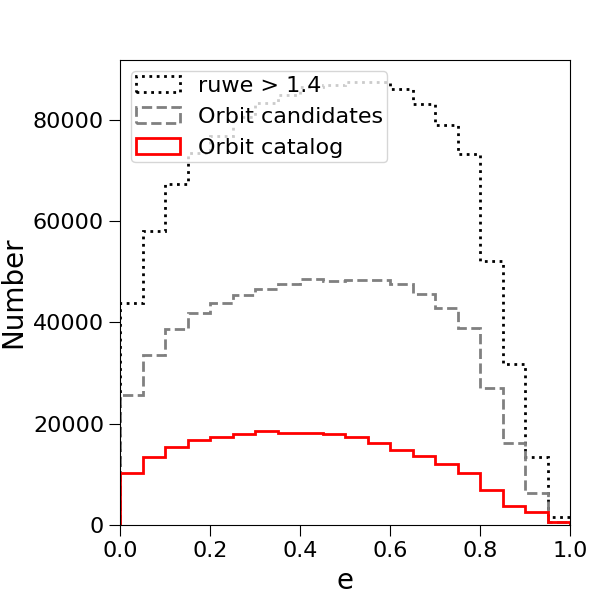}
    \caption{Effect of the different filters in the NSS processing cascade on the distribution of orbital periods (left) and eccentricity (right).
    Each line shows a different filter: $\ruwe > 1.4$ (black dotted), orbit candidates (gray dashed), and the final orbit catalog (red solid).
    The filters in the acceleration cascade preferentially remove long period orbital solutions.
    In the orbital cascade, the filter on $\varpi/\sigma_\varpi > 20,000 \textrm{ d}/P_{orb}$ preferentially removes short period and 1-year period solutions, while the filter on $a_0/\sigma_{a_0} > \max(5, 158/ \sqrt{P_{orb}/\textrm{days}})$ also preferentially removes eccentric solutions.
    \label{fig:fm_cuts_period_ecc}
    }
\end{figure*}

Figure \ref{fig:fm_corner} is a corner plot comparing the orbital period, parallax, eccentricity, photocenter orbit semimajor axis, \ruwe, apparent magnitude, and parallax over parallax error of the mock forward model to the actual NSS astrometric orbit catalog.
The 2-D correlations are preserved, and the 1-D distributions are also generally reasonable matches.
The orbital period distribution is a good match between the forward model and actual NSS catalog.
The distribution of parallaxes in the forward model are systematically offset larger than the distribution of NSS parallaxes by about 0.1 dex; otherwise, the agreement is also good.
The distribution of $a_0$ and $\ruwe$ are excellent matches.
The forward model apparent magnitude distribution is somewhat less peaked and has slightly more bright sources than the NSS distribution, but the match is otherwise good.
The distribution of parallax over error in the forward model are systematically offset larger than the distribution of NSS parallaxes by about 0.3 dex; part of this may be related to the parallaxes being high by 0.1 dex.
The eccentricity distribution of the model and the true catalog are the most discrepant parameter, with the \emph{Gaia} NSS distribution peaking around $e \approx 0.2$, while the forward modeled distribution is flatter and has a sharp drop-off in the number of solutions around $e \approx 0.6$;
we discuss possible reasons for this in \S \ref{sec:Discrepancy in eccentricity}.

\subsection{Cascade selection effects}

Next, we explore the selection effects of each part of the cascade and how they produce the resultant distributions in Figure \ref{fig:fm_corner}, in particular for the orbital period and eccentricity distributions.

Figure \ref{fig:alive_bright_cuts_2} compares the 1-D distribution of binary parameters in the initial birth population (dashed gray line) with that obtained from a simple filter on systems that could be astrometrically characterized by Gaia (dotted gray line), defined to be $P_{orb} < 1500$ days and $G < 19$ mag.
It also shows the distribution of parameters after application of the $\ruwe > 1.4$ filter (black line), removal of acceleration solutions (blue line), and the final NSS astrometric orbit catalog (red line).
Table \ref{tab:filter number} summarizes the effects of each filter on the population of sources.
We now discuss in detail the impact of each step in the NSS astrometric cascade.

\subsubsection{Input catalog}

The initial selection of NSS astrometric candidates primarily introduces a selection effect on the orbital period of the solutions.
The $\ruwe > 1.4$ filter preferentially selects sources with $P_{orb} \approx 1000$ days and the orbital period distribution is roughly symmetric in $\log_{10}(P_{orb})$
(Figure \ref{fig:alive_bright_cuts_2}).
As previously mentioned, our $\ruwe$ model is only accurate for $P_{orb} < 1500$ days, so we validated that the decrease in number of solutions held out to $P_{orb} = 10,000$ days using the catalog of \cite{Kareem}.
Orbits with $P_{orb} \approx 10^{3 \pm 1}$ days are the sweet spot in the $\ruwe$ criteria: those with $P_{orb} < 100$ days will be too small to have a significant photocenter wobble given the measurement uncertainties; those with $P_{orb} > 10,000$ days will lack significant orbital coverage during Gaia DR3 and can be approximately modeled by a single-star fit, given the measurement uncertainties.
An orbit of 1000 days maximizes the size of the photocenter wobble and orbital coverage during the survey.

Next, we consider the effects on the eccentricity distribution (Figure \ref{fig:alive_bright_cuts_2}).
Short period orbits are preferentially removed by the \ruwe\, filter, and most short period orbits $P_{orb} \lesssim 30$ days are circular \citep{Moe:2017}.
Although \ruwe\, also directly removes high eccentricity orbits (Equation \ref{eq:ruwe_G_a0_e_i}), it is a weak effect.

Finally, the $\ruwe$ filter also preferentially selects nearby sources with large photocenters and photocenter scales (Figure \ref{fig:alive_bright_cuts_2}).

\subsubsection{Acceleration models}

Next, we analyze the effect of the acceleration models.
Recall that sources that have an accepted acceleration model during the main NSS processing are \emph{not} able to be orbital solution candidates.

The acceleration portion of the NSS cascade preferentially removes solutions with $P_{orb} > 500$ days (Figure \ref{fig:fm_cuts_period_ecc}).
Between $P_{orb} \approx 500 - 2000$ days, the fraction of NSS astrometric candidates that have accepted acceleration solutions increases from 0 to about 85\%.
Because long $P_{orb}$ solutions are more likely to only have partial orbital period coverage, the acceleration models are more likely to be accepted; orbits with full coverage are more likely to be poor fits to the acceleration models.
The acceleration portion of the NSS cascade explains why the peak of the orbital period distribution is at such a short $P_{orb} \approx 500$ days.
In the absence of the acceleration cascade, one would expect high sensitivity to $P_{orb} \approx 700 - 900$ day sources given the $\approx 1000$ day time baseline of Gaia DR3.

The removal of acceleration solutions does not substantially change the eccentricity distribution (Figure \ref{fig:fm_cuts_period_ecc}).

\subsubsection{Orbital model selection criteria}

Finally, the last step is selecting the orbital model candidates that will actually be accepted into the final catalog.
The orbital model significance filter $a_0/\sigma_{a_0} >5$ has a very strong effect on the eccentricity distribution, preferentially removing eccentric solutions $e \gtrsim 0.6$.
It also preferentially removes long-period solutions $P_{orb} \gtrsim 1000$ days.
This is because at a fixed source magnitude and number of visibility periods, solutions which do not have good orbital coverage will have larger uncertainties on $a_0$.
Similarly, eccentric solutions will also have larger uncertainties on $a_0$.
The $a_0/\sigma_{a_0} > 158/\sqrt{P_{orb}/\textrm{d}}$ filter also preferentially removes eccentric systems, although it preferentially removes short period systems as well because of the $1/\sqrt{P_{orb}}$ dependence.

The parallax significance relative to orbital period filter $\varpi/\sigma_\varpi > 20,000 \textrm{d}/P_{orb}$ does not bias against the eccentricity.
It preferentially removes short period solutions due to the $1/P_{orb}$ dependence.
It also preferentially removes solutions with $P_{orb} \approx 1$ year because these sources have a very low $\varpi/\sigma_\varpi$, due to the large errors from the degeneracy between parallax and orbital motion.
Although this effect can be seen in the previous two filtering steps, it is much stronger in this step.

\subsection{Probability of finding \emph{Gaia} black holes 
\label{sec:Probability of finding Gaia BHs}}

\begin{deluxetable*}{c|cccccccccc|c}
\tablecaption{\emph{Gaia} black hole properties and detection probabilities}
\label{tab:gaia bh properties}
\tablehead{
    \colhead{Name} &
    \colhead{RA (deg)} & 
    \colhead{Dec (deg)} & 
    \colhead{$P_{orb}$ (day)} & 
    \colhead{$e$} &
    \colhead{$\varpi$ (mas)} &
    \colhead{$M_* (M_\odot)$} &
    \colhead{$M_\bullet (M_\odot)$} &
    \colhead{$G_*$ (mag)} &
    \colhead{$M_{G,0*}$ (mag)} &
    \colhead{$A_G$ (mag)} & 
    \colhead{p(NSS)}}
\startdata
Gaia BH1 & 262.17 & --0.58 & 185 & 0.43 & 2.09 & 0.93 & 9.24 & 13.77 & 4.6 & 0.76 & 0.38 \\
Gaia BH2 & 207.57 & --59.24 & 1277 & 0.52 & 0.859 & 1.07 & 8.94 & 12.28 & 1.8 & 0.15 & 0.27 \\
\enddata
\tablecomments{
Properties of Gaia BH1 and BH2.
$*$ subscript denotes properties of the stellar companion, $\bullet$ subscript denotes properties of the black hole.
All values except $M_{G,*}$ and $A_G$ are taken from \cite{Nagarajan:2024} and \cite{El-Badry:2023b} for Gaia BH1 and BH2, respectively.
We note for Gaia BH2 that our extinction map suggests much less extinction ($\approx 0.4$ mag) than the map used by \cite{El-Badry:2023b} and so our inferred extinction-corrected absolute magnitude of the star $M_{G,0*}$ in Gaia BH2 is fainter.
We do not calculate the probability for Gaia BH3 \cite{Panuzzo:2024} because its orbital period is longer than 1500 days.}
\end{deluxetable*}

We apply our forward modeling framework to determine the probability of inclusion of \emph{Gaia} BH1 \citep{El-Badry:2023a, Chakrabarti:2023} and \emph{Gaia} BH2 \citep{El-Badry:2023b} in the \emph{Gaia} DR3 NSS astrometric orbits catalog.
We do not calculate the probability for \emph{Gaia} BH3 \citep{Panuzzo:2024} because its orbital period is much longer than 1500 days, where our \ruwe\, model breaks down.
We calculate the probability as a function of sky position, orbital period, eccentricity, parallax, and the component magnitudes and masses. 
We integrate over random inclinations that are uniform in $|\cos i|$.

\begin{figure*} %[t]
    \centering
    \includegraphics[width=0.495\linewidth]{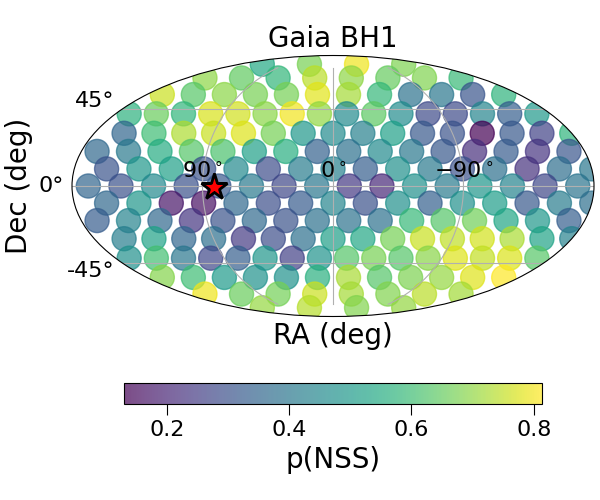}
    \includegraphics[width=0.495\linewidth]{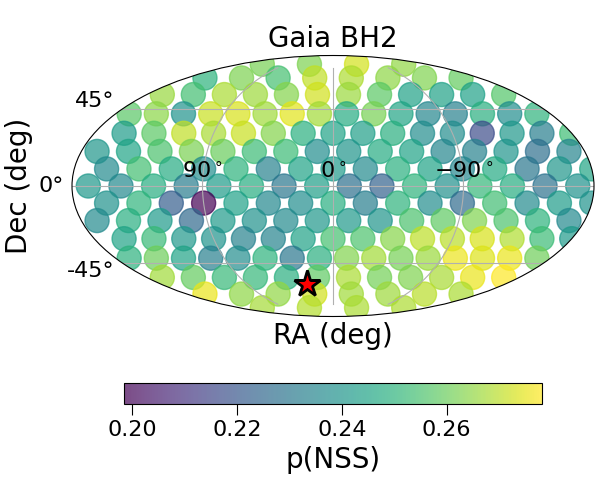}
    \includegraphics[width=0.495\linewidth]{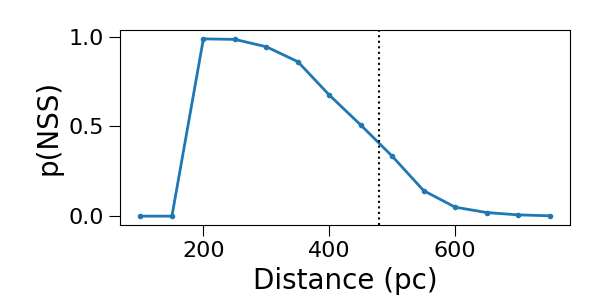}
    \includegraphics[width=0.495\linewidth]{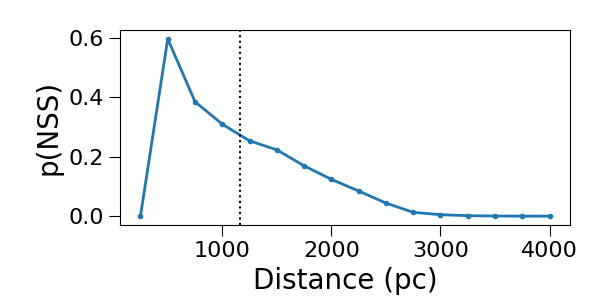}
    \caption{
    \emph{Top panels}: Probability of inclusion of Gaia BH1 (\emph{left)} and Gaia BH2 (\emph{right}) in \emph{Gaia} DR3 NSS astrometric orbits catalog as a function of sky position (192 equally spaced points on sky); the color of the circle denotes the probability of inclusion.
    Note that the color bar scales in the two panels are different. 
    We fix all parameters as listed in Table \ref{tab:gaia bh properties}, and only vary the sky position.
    We sample randomly over inclination $|\cos i|$.
    We calculate the extinction using \texttt{mwdust}.
    The positions at which Gaia BH1 and BH2 were detected are marked as the red stars.
    For Gaia BH1, the detectability as a function of sky position clearly shows the pattern of the scanning law.
    The ecliptic plane receives fewer scans than the ecliptic poles \citep{Lindegren:2012}, reducing the astrometric precision and in turn the size of a detectable astrometric orbit.
    Gaia BH1 thus tends to be more reliably detected around the ecliptic poles than the ecliptic plane.
    For Gaia BH2, the detectability does not have as strong a dependence on the sky position.
    \emph{Bottom panel}: Probability of inclusion of Gaia BH1 (\emph{left)} and Gaia BH2 (\emph{right}) in \emph{Gaia} DR3 NSS astrometric orbits catalog as a function of the system's distance.
    We fix all parameters as listed in Table \ref{tab:gaia bh properties}, and only vary the distance of the system.
    We sample randomly over inclination $|\cos i|$.
    We calculate the extinction using \texttt{mwdust}.
    The distances at which Gaia BH1 and Gaia BH2 were detected at are denoted by the dotted black lines.
    \label{fig:bh_sky_distance}
    }
\end{figure*}

We find the probability of Gaia BH1 ((RA, Dec) = (262.17, --0.58) deg, $P_{orb} = 186$ day, $e = 0.45$, $\varpi = 2.09$ mas, $M_* = 0.93 M_\odot$, $M_\bullet = 9.62 M_\odot$, $G_* = 13.77$ mag) being included in the \emph{Gaia} DR3 NSS astrometric orbit catalog to be 38\%.
We also vary the location of BH1 on the sky and its distance from the Sun (Figure \ref{fig:bh_sky_distance}) and consider how its probability of inclusion in the \emph{Gaia} DR3 NSS astrometric orbit catalog changes. 
Since Gaia BH1 is located near the ecliptic plane, it had a relatively low probability of inclusion; toward the ecliptic poles, the inclusion probability would have been closer to 60--80\%.
Gaia BH1 is also relatively close by, at a distance of about 500 pc. 
Had it been at the same (RA, Dec) but slightly farther away, at 600 pc or more, it would not have made it into the NSS catalog.
Conversely, had it been slightly closer, 400 pc or less, it would have had $>80\%$ probability of making it into the catalog.
This suggests that if there are Gaia BH1-like systems within a few hundred pc of the Sun, they should have been included in the \emph{Gaia} DR3 NSS astrometric orbits catalog; the lack of such detections suggests that there are no or few Gaia BH1-like sources within 300--400 pc of the Sun.

We find the probability of Gaia BH2 ((RA, Dec) = (207.57, --59.24), $P_{orb} = 1277$ day, $e = 0.52$, $\varpi = 0.86$ mas, $M_* = 1.07 M_\odot$, $M_\bullet = 8.94 M_\odot$, $G_* = 12.28$ mag) being included in the \emph{Gaia} DR3 NSS astrometric orbit catalog to be 27\%.
We also vary the location of BH2 on the sky and its distance from the Sun (Figure \ref{fig:bh_sky_distance}) and consider how its probability of inclusion in the \emph{Gaia} DR3 NSS astrometric orbit catalog varies. 
Since Gaia BH2 has a long orbital period, the main factor of whether it made it into the NSS catalog is whether or not it was assigned an acceleration solution or not.
% The inclusion probability of Gaia BH2 does not change significantly with distance until beyond $\gtrsim 2.5$ kpc, at which point the inclusion probability substantially drops.
The inclusion probability shows a weak dependence on ecliptic latitude, and does not vary significantly ($<10$\%) across the sky.
Gaia BH2 is around 1.2 kpc from the Sun.
Had it been at the same (RA, Dec) but beyond 3 kpc, it would not have been included in the NSS catalog.
Conversely, had it been closer (between 500 and 1000 pc), it would have a somewhat higher probability ($50-60$\%) of making it into the NSS catalog.

The results here for BH1 are consistent with the results in \cite{Kareem}, which find a 39\% probability of inclusion in the Gaia DR3 NSS astrometric orbit catalog.
The probability of inclusion for BH1 as a function of distance (Figure 15 of \citealt{Kareem}) is also very similar.
For BH2, \cite{Kareem} finds a higher probability of 48\%.
The probability of inclusion for BH2 as a function of distance (Figure 15 of \citealt{Kareem}) is also similar, albeit their inclusion probabilities are generally about $10-20$\% higher.
Since our model was tuned to match that of \cite{Kareem}, we might expect the results to be more similar.
This may be due to the probabilistic nature of the acceleration cascade modeling; $P_{orb}$ and $a_0/\sigma_{AL}$ may not capture the full dependencies of probability acceptance, leading to large scatter.
This suggests the methodology may be reliable only to a factor of ${\sim}2$ beyond $P_{orb} \gtrsim 500$ days.

\subsection{Inferences on the population of Galactic detached black holes}

\begin{figure*}
    \centering
    \includegraphics[width=0.49 \linewidth]{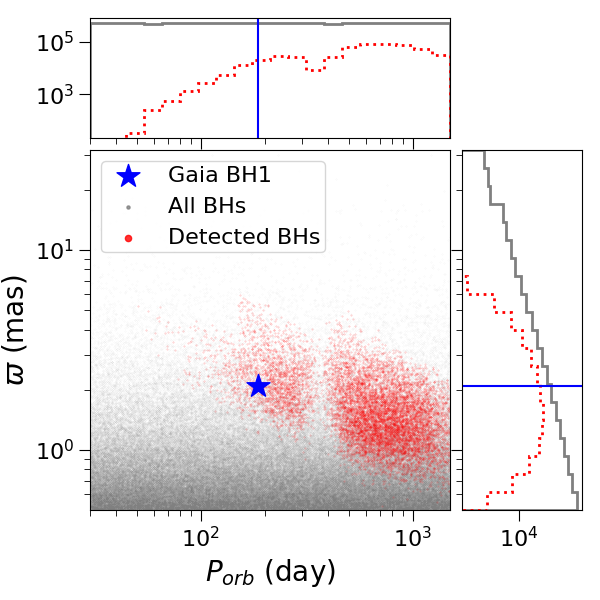}
    \includegraphics[width=0.49 \linewidth]{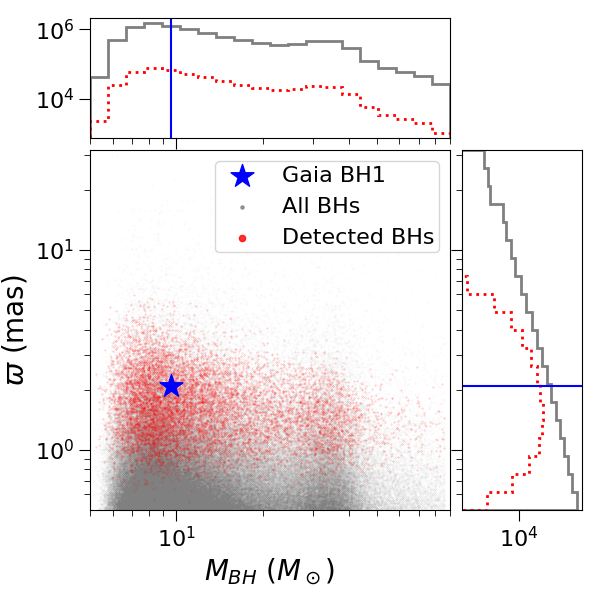}
    \includegraphics[width=0.49 \linewidth]{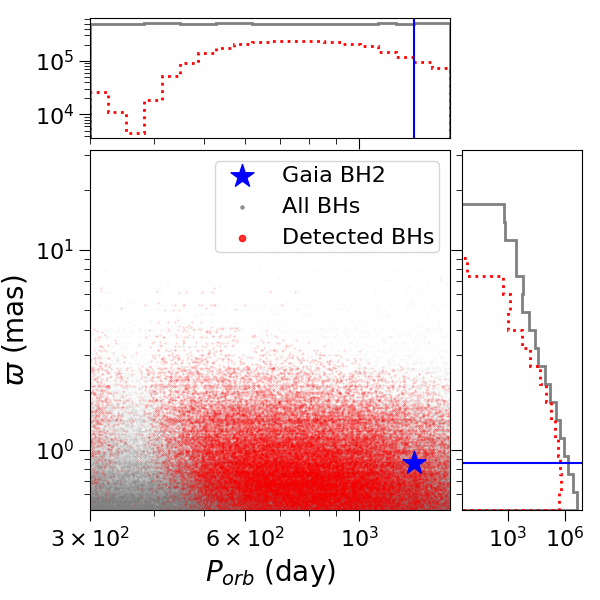}
    \includegraphics[width=0.49 \linewidth]{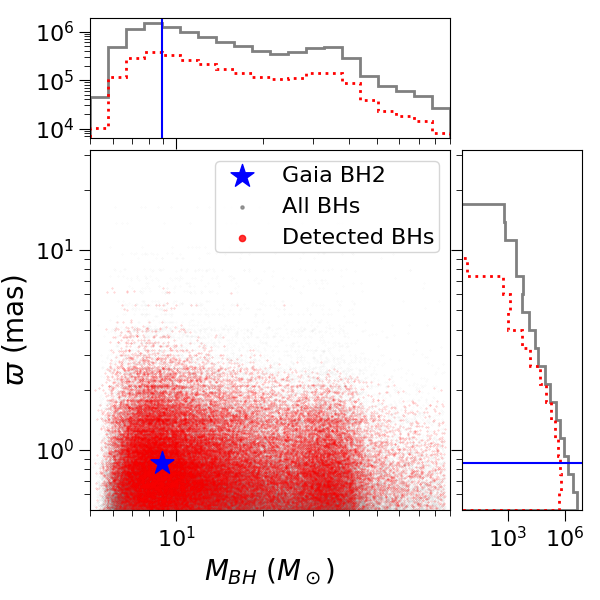}
    \caption{
    Parallax vs. orbital period and black hole mass for a simulated population of detached Galactic BHs within 2 kpc (\emph{top panels:} Sun-like stars + BHs, \emph{bottom panels:} red giants + BHs).
    The properties of Gaia BH1 and Gaia BH2 are marked in the top and bottom panels, respectively.
    The detectability as a function of orbital period resembles the NSS period distribution.
    For Sun-like stars + BHs, most are detectable with $\varpi > 1$ mas, but red giants + BHs can be detected with $\varpi > 0.5$ mas and lower.
    Detectability does not strongly depend on the BH mass.
    \label{fig:porb_bhmass_parallax}
    }
\end{figure*}

As an application of the forward model and selection function, we calculate the detection probabilities of systems like Gaia BH1 and BH2 in the Milky Way.

First, we generate a synthetic population of detached BH + star binaries.
We assume the distribution of black hole masses follows the primary mass distribution inferred by gravitational wave measurements.
We use the ``Power law + peak" model \citep{Talbot:2018}, a normalized power-law distribution with negative spectral index (i.e., more low-mass BHs than high-mass BHs) with an additional Gaussian peak.
The parameters for the model are inferred from the 2nd Gravitational Wave Transient Catalog \citep[GWTC-2;][]{Abbott:2021}, resulting in a mass distribution with a primary peak at $8 M_\odot$, with a smaller excess at $33 M_\odot$, notably consistent with the masses of the three Gaia BHs.
The orbital period distribution is assumed to be logarithmically flat \citep[e.g.,][]{MarinPina:2024}.
The eccentricity distribution is also assumed to be flat.
From \texttt{galaxia} we sample the positions (RA, Dec, parallax) of Sun-like stars for Gaia BH1, and red giants for Gaia BH2.
We use the \texttt{mwdust} 3-D dust map as described earlier to apply extinction.

For Gaia BH1-like sources, we assume a $1 M_\odot$ star with absolute magnitude $M_{G,0} = +4.6$.
We consider orbital periods $30 < P_{orb} < 1500$ days, spanning roughly the range of detected NSS astrometric orbits.
We will refer to these as Sun-like star + BH binaries, or Sun+BH binaries for short.

For Gaia BH2-like sources, we assume a $1 M_\odot$ star with absolute magnitude $M_{G,0} = +1.8$.
We consider orbital periods $300 < P_{orb} < 1500$ days.
The minimum orbital period is 300 days such that the semimajor axis of the orbit is larger than the typical red giant radius.
We will refer to these as red giant + BH binaries, or RG+BH binaries for short.

For the Sun+BH and RG+BH sources, we randomly pair positions and parallaxes with orbital periods, eccentricities, and black hole masses. 
Thus, there is no correlation between orbital period, parallax, BH mass, or position.
We randomly sample 10 million of these pairs, then use the selection function to determine whether or not each object would have been included in the \emph{Gaia} DR3 NSS astrometric orbits catalog.

Figure \ref{fig:porb_bhmass_parallax} shows the orbital periods, parallaxes, and black hole masses for the full population vs. detected systems. 
On average, 5\% of Sun+BH binaries with orbital periods $30 < P_{orb} < 1500$ days within 2 kpc are \emph{Gaia} DR3 detectable, and 27\% of RG+BH binaries with orbital periods $300 < P_{orb} < 1500$ days within 2 kpc are \emph{Gaia} DR3 detectable\footnote{On average, 10\% of Sun+BH binaries with orbital periods $300 < P_{orb} < 1500$ days within 2 kpc are \emph{Gaia} DR3 detectable.}.
The period distribution unsurprisingly looks similar to the NSS astrometric orbits catalog.
Sun+BH systems are generally difficult to detect beyond 1 kpc.
RG+BH systems can be detected to at least 2 kpc because the red giant is much more luminous than a Sun-like star despite having the same mass.
There is no strong dependence of the detectability of the system on black hole mass within the $5 - 80 M_\odot$ range.
The size of the photocenter orbit semimajor axis $a_0$ (Equation \ref{eq:photocenter}) increases by a factor of only 2.4 for a $5 M_\odot$ vs. $50 M_\odot$ black hole and $1 M_\odot$ star, even though the black hole mass increases by a factor of 10.

We can use this model to make a simple argument on the occurrence rate of detached black holes in the Milky Way in various orbital period ranges.
We first consider Sun+BH binaries with orbital periods $100 < P_{orb} < 400$ days, $400 < P_{orb} < 1000$ days, and $1000 < P_{orb} < 1500$ days, and separately RG+BH binaries with orbital periods $400 < P_{orb} < 1000$ days and $1000 < P_{orb} < 1500$ days.
We assume that Gaia BH1 (Sun+BH with an orbital period $100 < P_{orb} < 400$ days) and BH2 (RG+BH with an orbital period $1000 < P_{orb} < 1500$ days) are the only two star + BH binaries in the \emph{Gaia} DR3 astrometric orbits catalog, as the studies that have identified candidates in the astrometric NSS catalog \citep{Andrews:2022, Shahaf:2023} have been followed up and not yielded detections. Additional candidates identified in the Gaia BH1 discovery papers \citep{El-Badry:2023a, Chakrabarti:2023} also have been ruled out.

We then use our model to calculate the number of detectable detached BHs within 2 kpc of the Sun within these period ranges.
For the period ranges where no BH has been detected, we take this to be the upper limit on the number of those BH binaries.
Assuming there are 100 billion stars in the Galaxy, 1 billion, or 1\% of those stars are within 2 kpc of the Sun; to convert our number of objects within 2 kpc to the number of objects in the Galaxy, we multiply by 100.
% 160 million stars within 1 kpc, which is 0.1\% of all the stars in the Milky Way. 
% NOTE: Galaxia (before rescaling): 246953200 stars within 1 kpc, 1132486600 stars within 2 kpc.

Table \ref{tab:n_bhs} lists the detection probabilities. 
We find that the Galaxy hosts $3000 \pm 500$ Sun+BH binaries with $100 < P_{orb} < 400$ days, and $400 \pm 200$ RG+BH binaries with $1000 < P_{orb} < 1500$ days.
We also find that there are $<800$ Sun+BH binaries with $400 < P_{orb} < 1000$ day, $<12,000$ Sun+BH binaries with $1000 < P_{orb} < 1500$ day, and $<300$ RG+BH binaries with $400 < P_{orb} < 1000$ day.
These uncertainties are assuming Poisson statistics; they do not account for systematic errors (e.g., errors in the forward model, or the assumed number of stars in the Galaxy.)

Although the previous paragraphs treats the constraints on Sun+BHs separately from RG+BHs separately, in reality these constraints are related, as Sun-like stars will eventually evolve into red giants.
Because the evolutionary timescale of the red giant phase is only a $\sim 10\%$ of a star's main sequence lifetime, the number of RG+BHs should also be a few percent of the number of Sun+BH binaries in a given period range.
This is generally consistent with the results we find.
For the $400 < P_{orb} < 1000$ day range, we only have upper limits for RG+BH and Sun+BH binaries and so this is not particularly constraining.
For the $1000 < P_{orb} < 1500$ day range, this implies that the number of RG+BH binaries are likely on the lower side of the estimated range, and for the Sun+BH binaries, the number of systems will be close to the upper limit and they just barely managed to escape inclusion in the NSS astrometric orbits catalog.

In comparison, \cite{El-Badry:2023a} estimated $\sim 40,000$ low mass star + BH binaries with $100 < P_{orb} < 300$ days with the Galaxy, and an upper limit of 4,000 binaries for $400 < P_{orb} < 1000$ days, also assuming a logarithmically flat orbital period distribution.
We suspect the difference is due to how \cite{El-Badry:2023a} estimated $\varpi/\sigma_\varpi$ and $a_0/\sigma_{a_0}$ to model the \emph{Gaia} DR3 selection.
Because of this, their effective search volume is smaller, leading to a higher occurrence rate than in this work.

\begin{deluxetable}{ccc}
\tablecaption{Number of black holes}
\label{tab:n_bhs}
\tablehead{
    \colhead{$P_{orb}$ range (day)} & 
    \colhead{Prob. detect} & 
    \colhead{N($d < 2$ kpc)}}
\startdata
\multicolumn{3}{l}{Sun-like star + BH} \\
\hline
$100 < P_{orb} < 400 $ & 0.03 & $30 \pm 5$ \\
$400 < P_{orb} < 1000 $  & 0.14 & $<8$ \\
$1000 < P_{orb} < 1500 $  & 0.09 & $<12$ \\
\hline
\hline
\multicolumn{3}{l}{Red giant + BH} \\
\hline
$400 < P_{orb} < 1000 $  & 0.36 & $<3$ \\
$1000 < P_{orb} < 1500 $  & 0.25 & $4 \pm 2$ \\
\enddata
\tablecomments{
Number of black holes within 2 kpc of the Sun.
Upper limits and uncertainties are $1\sigma$, assuming Poisson errors.
To convert these results to the number of black holes in the entire Milky Way, multiply the values in the N($d < 2$ kpc) column by 100.}
\end{deluxetable}

\section{Discussion 
\label{sec:Discussion}}

\subsection{Discrepancy in eccentricity
\label{sec:Discrepancy in eccentricity}}

There is a discrepancy between the \emph{Gaia} NSS astrometric orbit and forward modeled eccentricity distribution, with the forward model predicting significantly more eccentric binaries than are present in the \emph{Gaia} astrometric orbits catalog.
We hypothesize that the starting eccentricity distribution used in the forward model is incorrect and leads to this discrepancy.

The period, eccentricity and mass ratio distributions are generally well known for short period binary systems with $P_{\rm orb} \approx 0.1 - 50$ days \citep[e.g.,][]{Moe:2015, Soszynski:2016, Rowan:2022}.
At longer orbital periods, the primary search techniques of radial velocities and astrometry are both biased against the detection and characterization of high eccentricity binaries.  
The $P_{\rm orb}-e$ relationship for $P_{\rm orb} \gtrsim 50$ day is generally modeled by extrapolation from a small number of moderate eccentricity systems, and is therefore poorly known and highly sensitive to the assumed completeness correction \citep{Moe:2017}.
Because of these systematic biases, the intrinsic eccentricity distribution for wide binaries is highly uncertain. 

Over time, the eccentricity distribution for Sun-like stars with $10 \lesssim P_{orb} \lesssim 1000$ days has been inferred to be more and more eccentric.
\cite{Duquennoy:1991} found a Gaussian eccentricity distribution with mean value $e \approx 0.3$.
An updated study by \cite{Raghavan:2010} instead found evidence for a flat eccentricity distribution.
\cite{Moe:2017} used the \cite{Raghavan:2010} data to infer a power-law distribution in eccentricity $p(e) \propto e^\eta$, with $\eta = 0.2 \pm 0.3$ for $100 < P_{orb} < 1000$ days. 
\cite{Moe:2017} then combined their result with other literature datasets and found for main sequence binaries with primary masses $0.8 M_\odot < M_1 < 3 M_\odot$ and orbital periods $0.5 < \log P_{orb} < 6.0$ days,
\begin{equation}
    \eta = 0.6 - \frac{0.7}{\log P_{orb}/\textrm{day} - 0.5}
    \label{eq:eccentricity}
\end{equation}
(Equation 17 of \citealt{Moe:2017}).
This parametrization results in $\eta = 0.13$ for $P_{orb} = 100$ days and $\eta = 0.32$ for $P_{orb} = 1000$ days, and was the assumed eccentricity distribution used in the forward model presented in this work.

\begin{figure*}
    \centering
    \includegraphics[width=\linewidth]{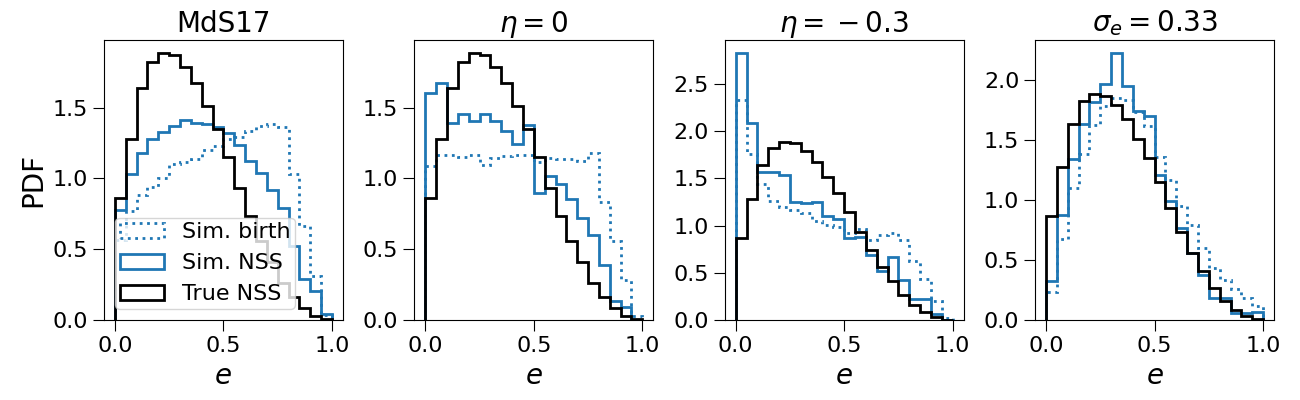}
    \caption{Forward model assuming different birth eccentricity distributions.
    The birth distribution of eccentricities for orbital periods within $30 < P_{orb} < 1500$ days is shown in the dotted blue line.
    The forward modeled eccentricity distribution is shown in the solid blue line.
    The distribution of eccentricities in the NSS astrometric orbits catalog is shown in black.
    The default prescription in \cite{Moe:2017} (left panel) produces an excess of high-eccentricity orbits.
    Assuming a flat eccentricity distribution (left middle panel) $\eta = 0$ is a slight improvement.
    Assuming a distribution with $\eta = -0.3$ (right middle panel) matches the long-eccentricity tail well, but overestimates the number of circular orbits.
    A Rayleigh distribution with $\sigma_e = 0.33$ (right panel) provides the best match to the observed eccentricity distribution.
    \label{fig:eta}}
\end{figure*}

As a test of our forward model to see if a different eccentricity distribution would result in a better match to the observed NSS distribution, we re-run the population synthesis, replacing Equation \ref{eq:eccentricity} from \cite{Moe:2017} with $\eta = 0$ and $-0.3$.
All other parts of the population synthesis remain the same.
We then re-run the forward model and look at the resulting eccentricity distribution.

Figure \ref{fig:eta} compares the different eccentricity prescriptions.
Although neither of them provides a great match, they are improvements at the high eccentricity end.
The issue with a power law parametrization is that any decrease in high-eccentricity solutions is compensated by more circular solutions.
Therefore, a different functional form is likely needed.
The observed NSS eccentricity distribution peaks at $e\approx 0.2 - 0.3$.
Qualitatively, the birth distribution needs to have a concentration of solutions around $e = 0.2 - 0.3$ to match the observed \emph{Gaia} NSS distribution.

\cite{Wu:2025} propose a Rayleigh distribution
\begin{equation}
    p(e) = \frac{e}{\sigma_e^2} \exp \left[ -\frac{e^2}{2\sigma_e^2} \right]
    \label{eq:rayleigh}
\end{equation}
to model the eccentricity distribution of \emph{Gaia} DR3 NSS sources. 
They find that for their ``gold sample", which consists of $\sim 3000$ Sun-like binaries with $10^2 < P_{orb} < 10^3$ days within 150 pc, the best-fit scale parameter is $\sigma_e = 0.336$, which includes a completeness correction to account for NSS selection effects.
We test this using our forward model.
For our original synthetic population, for any source with $\log_{10} P_{orb} > 0.7$, we instead draw their eccentricities from Equation \ref{eq:rayleigh} with $\sigma_e = 0.33$.
Because the Rayleigh distribution has support over $[0, \infty)$, we remove any sources that have $e > 1$.
Following \cite{Moe:2017}, we also removed sources where the eccentricity exceeded 
\begin{equation}
    e_{max}(P_{orb}) = 1 - \left( \frac{P_{orb}}{2 \textrm{ d}} \right)^{-2/3}
\end{equation}
to ensure that the components have Roche lobe fill factors $\lesssim70\%$ at periastron.
We simply enforce a hard cutoff at $e_{max}$; in contrast, COSMIC applies a linear turnover in the number of eccentric solutions at $e / e_{max} = 0.8$, such that the eccentricity distribution is continuous at $e / e_{max} = 0.8$ and zero at $e = e_{max}$.
Finally, we again run our NSS forward model to determine the resulting eccentricity distribution (rightmost panel of Figure \ref{fig:eta}), which now closely reproduces the observed NSS distribution.
As noted in \cite{Wu:2025}, the birth and NSS forward modeled eccentricity distributions are very similar, meaning the eccentricity distribution is approximately complete.
This is because a Rayleigh distribution with $\sigma_e \approx 0.3$ has very few high-eccentricity solutions, where the distribution is most affected by selection effects; it is not generally true of any eccentricity distribution.

With \emph{Gaia} DR4 per-epoch astrometry, these findings can be tested and validated.
As a cross check, we also advocate for an independent method of measuring eccentricities that is less biased against high-eccentricity systems.
Measuring the eccentricities of long-period eclipsing binaries from the \emph{Kepler} survey \citep{Kirk:2016} is one way to accomplish this goal.
\emph{Kepler} continually monitored a single part of the sky over 4 years and is thus sensitive to eclipsing binaries with orbital periods of up to $\approx 2$ years.

\subsection{Strengths and limitations of the approach
\label{sec:Strengths and limitations of this forward model approach}}

In the spirit of keeping calculations fast and analytic in this work, we construct empirical probabilistic functions to approximate the fitting of epoch 1-D astrometry.
This has both strengths and limitations, namely the tradeoff between speed and simplicity vs. computational expense and accuracy.

The main strength of a fast simulation is for exploring parameter space.
It requires approximately $10-20$ seconds to run the full astrometric cascade of \cite{Kareem} for each source that receives an orbital solution, or about 5000 CPU hours to produce a full mock DR3 binary catalog.
In contrast, it requires approximately 2 seconds to run 100,000 evaluations of the empirical forward model of this work at a fixed position on sky; most of that time comes from querying the 3-D dust map.
Of course, it is important to note that the empirical corrections to the analytic prescription are dependent on having the expensive epoch modeling simulations already in hand.
It is analogous to significant upfront costs in training machine learning models, but later having substantially lower costs when using the model to make predictions.
Finally, the simple approach also allows for a more intuitive understanding of each step in the cascade and elucidates the complex selection function.

The speed and simplification comes at the expense of detail and potentially accuracy.
The major limitation of our analytic models is that they are limited to certain regions of parameter space.
Most notably, as discussed earlier, our model does not cover orbital periods $P_{orb} > 1500$ days.
In addition, when trying to understand one specific system in detail, individual epoch modeling will likely be more accurate and can provide specific insight.
One way to take advantage of the strengths of both the fast and detailed approaches are to use the fast simulations to identify potentially interesting regions of parameter space, then to use the detailed simulations to validate and refine those results.

\subsection{Spectroscopic solutions}

\cite{Halbwachs:2023} discuss in their \S 5.3 ``Selection of the orbital solutions" that 2470 astrometric orbit solutions were recovered after being rejected by the post-processing by combining them with single-lined spectroscopic binary (SB1) solutions, and another 3500 solutions were discarded after post-processing due to validation described in \cite{Babusiaux:2023}.
The processing description for the SB1 solutions was recently published \citep{Gosset:2024}.
These solutions represent about 3.5\% of the total number of solutions in the NSS catalog, and so are likely negligible at the accuracy of the forward model presented in this work.
However, it would be a limiting factor in achieving a precise understanding of selection effects and completeness; future work will incorporate the description of the SB1 processing.
Ultimately, the epoch data of \emph{Gaia} DR4 will mitigate this issue (\S \ref{sec:Gaia DR4}).

\subsection{\emph{Gaia} DR4
\label{sec:Gaia DR4}}

As of 2025, \emph{Gaia} DR4 is projected to be released in late 2026. 
DR4 will nearly double the time baseline of DR3 (66 months of data vs. 34 months). 
It is expected that the parallax and photocenter orbit semimajor axis uncertainties will be reduced by a factor of $\sqrt{2}$ due to having double the number of observations, and proper motion errors will be reduced by a factor of $\sqrt{8}$, with a factor of 2 in the doubled baseline, and another factor of $\sqrt{2}$ for the doubled number of observations.

It is beyond the scope of this work to extend this forward model framework to forecast for \emph{Gaia} DR4.
In principle, this can by done by re-generating the catalogs of \cite{Kareem} using the 66 month DR4 baseline, and re-calibrating the empirical corrections here.

Finally, we also note that DR4 will be the first \emph{Gaia} data release to include individual epoch photometry and astrometry.
Thus, the completeness and selection function can be evaluated in other manners, such as injection and recovery tests.

\section{Conclusions
\label{sec:Conclusions}}

The \emph{Gaia} DR3 NSS catalog has dramatically increased the number of well-characterized binary systems, and enabled the first mass measurement of black holes in detached binary systems.
However, the selection function of this catalog is not well-characterized, and without knowledge of the selection function, discoveries of interesting systems are limited to individual curiosities, as their population statistics cannot be inferred. 

To address this, we create a fast, simple, and physically interpretable framework to forward model the selection function of the \emph{Gaia} DR3 NSS astrometric orbits catalog. 
Our method combines analytic prescriptions with detailed results from \citet{Kareem}.
Using a binary population synthesis model, we validate this framework.
We find the method produces reasonable agreement with the observed distribution of periods, parallaxes, photocenter sizes, \ruwe, magnitudes, and parallax errors; the one parameter that is substantially discrepant is the eccentricity; we suspect the assumed birth population of eccentricities is the culprit.
We find the proposed Rayleigh distribution of \cite{Wu:2025} to better explain the eccentricity distribution of \emph{Gaia} sources.

As an example application of this framework, we use it the model to calculate the probability of Gaia BH1 and BH2 being included in the NSS astrometric orbits catalog.
We find that both Gaia BH1 and BH2 had relatively high probabilities ($>20$\%) of being included in the Gaia DR3 NSS astrometric orbits catalog, and that there are a few thousand Sun-like star + BH binaries in the entire Galaxy.

Ultimately, we look forward to \emph{Gaia} DR4, where the time baseline will be doubled, and per-epoch astrometry will be released for all sources.
This will enable proper characterization of the selection function and the discovery of many more interesting binary systems.

\smallskip

The code to calculate the probability of a binary being included in \emph{Gaia} DR3 is available at \url{https://github.com/caseylam/dr3_nss_ast_orbits_selection}.

\smallskip 

% \begin{acknowledgments}
C.Y.L. acknowledges support from the Harrison and Carnegie Fellowships.
K. E.-B. is supported by NSF grant AST-2307232.
This work has made use of data from the European Space Agency (ESA) mission
{\it Gaia} (\url{https://www.cosmos.esa.int/gaia}), processed by the {\it Gaia}
Data Processing and Analysis Consortium (DPAC,
\url{https://www.cosmos.esa.int/web/gaia/dpac/consortium}). Funding for the DPAC
has been provided by national institutions, in particular the institutions
participating in the {\it Gaia} Multilateral Agreement.
This work has made use of NASA’s Astrophysics Data System.
% \end{acknowledgments}

\vspace{5mm}
\facilities{\emph{Gaia}}

\software{isochrones \citep{Morton:2015}, galaxia \citep{Sharma:2011}, astropy \citep{Astropy:2013, Astropy:2018,Astropy:2022}, Matplotlib \citep{Hunter:2007}, NumPy \citep{vdWalt:2011}, SciPy \citep{Scipy:2019}, healpy \citep{Zonca:2019}, mwdust \citep{Bovy:2019}, COSMIC \citep{Breivik:2020}}

\appendix

\section{Eccentricity error $\sigma_e$ \label{app:eccentricity error}}

Table 1 of \cite{Halbwachs:2023} lists the criteria for acceptance as an orbital solution as $F_2 < 25$, $\varpi/\sigma_\varpi > 20,000 \textrm{ day}/P_{orb}$, $a_0/\sigma_{a_0} > \max(5, 158/ \sqrt{P_{orb}/\textrm{days}})$, and $\sigma_e < 0.079 \ln{P_{orb}/\textrm{d}} - 0.244$.
As described in \S 5.3 ``Selection of the orbital solutions" of \cite{Halbwachs:2023}, the criteria $a_0/\sigma_{a_0} > 158/ \sqrt{P_{orb}/\textrm{days}}$ and $\sigma_e < 0.079 \ln{P_{orb}/\textrm{d}} - 0.244$ only remove an additional 7\% of solutions after the other three criteria were applied.
Here, we provide justification to ignore $\sigma_e < 0.079 \ln{P_{orb}/\textrm{d}} - 0.244$ in our forward modeling.

\begin{figure*}
    \centering
    \includegraphics[width=\linewidth]{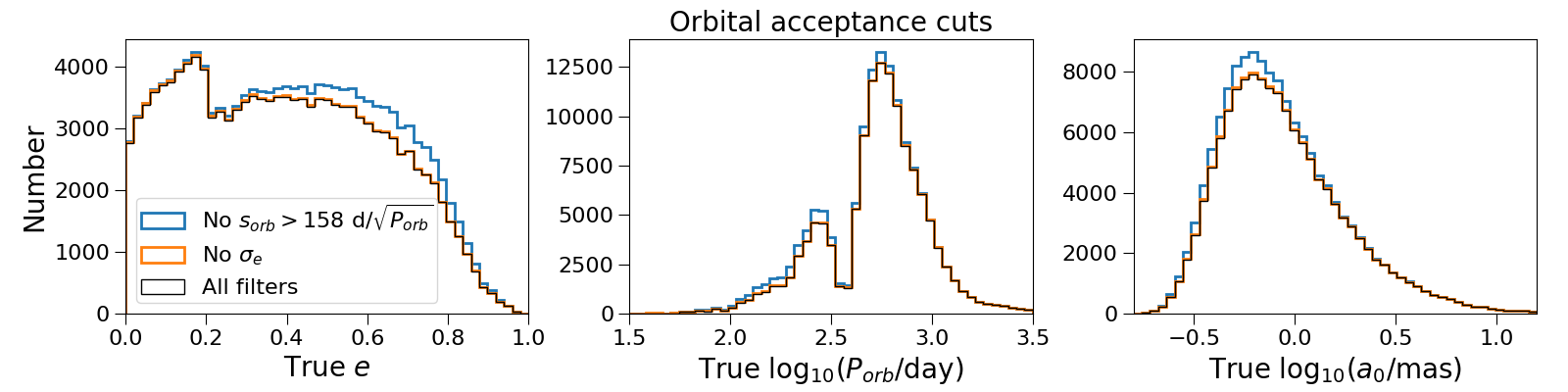}
    \caption{Distributions of the orbital period $P_{orb}$, eccentricity $e$, and photocenter orbit semimajor axis $a_0$ depending on whether the filters on $a_0/\sigma_{a_0} > 158/ \sqrt{P_{orb}/\textrm{days}}$ and $\sigma_e < 0.079 \ln{P_{orb}/\textrm{d}} - 0.244$ were included.
    \label{fig:ecc_err_cut}}
\end{figure*}

We analyze all the orbital solutions in the simulations of \cite{Kareem}  that pass the filters $\varpi/\sigma_\varpi > 20,000 \textrm{ day}/P_{orb}$ and $a_0/\sigma_{a_0} > 5$.
We then 1) remove solutions that also pass the filter $a_0/\sigma_{a_0} > 158/ \sqrt{P_{orb}/\textrm{days}}$, 2) remove solutions that pass the $\sigma_e < 0.079 \ln{P_{orb}/\textrm{d}} - 0.244$ filter, and 3) remove solutions that pass both $a_0/\sigma_{a_0} > 158/ \sqrt{P_{orb}/\textrm{days}}$ and $\sigma_e < 0.079 \ln{P_{orb}/\textrm{d}} - 0.244$ filters.
We show the distribution of $P_{orb}$, $e$, and $a_0$ for these solutions in Figure \ref{fig:ecc_err_cut}.
The effect of excluding the $\sigma_e < 0.079 \ln{P_{orb}/\textrm{d}} - 0.244$ filter makes almost no difference on the inferred parameter distributions.

\section{Goodness-of-fit $F_2$ \label{app:Goodness-of-fit F2}}

\begin{figure*}
    \centering
    \includegraphics[width=\linewidth]{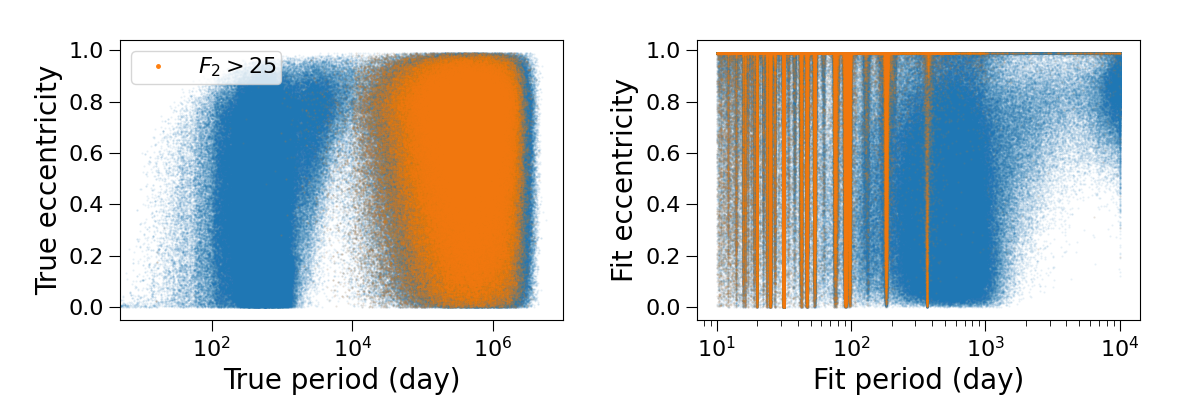}
    \caption{Spurious solutions as identified in the epoch modeling of \cite{Kareem}.
    Sources with $F_2 > 25$ (orange) tend to have incorrectly measured periods. 
    The inferred periods (right panel) are much shorter than the true periods ($P_{orb} \gtrsim 10,000$ days), which \emph{Gaia} DR3 is insensitive to.
    \label{fig:f2_cut}}
\end{figure*}

The $F_2 > 25$ criterion is mainly for removing spurious solutions. 
Because our modeling assumes the true orbital parameters are known and that all systems are binaries (and not, e.g., triple systems), these spurious solutions do not exist in our forward model. 
We validate this using the simulations in \cite{Kareem}. 
Figure \ref{fig:f2_cut} shows period vs. eccentricity for sources where an orbital solution is tried, with bad solutions ($F_2>25$) in orange points. 
The left panel shows the true period and eccentricity (which are known because they are simulation inputs).
The right panel are the inferred values performed from fitting simulated Gaia data (c.f. the middle panel of Figure 9 of \citealt{Kareem}). 
The solutions removed by the $F_2>25$ criterion are all artifacts due to incorrect fits and have true orbital periods $P_{orb} > 10,000$ days.
Given that we only model sources with $P_{orb} < 1500$ days, we are justified in assuming that $F_2 < 25$ is always satisfied for our sources.

\section{Acceleration modeling and orbital properties
\label{app:Acceleration modeling and orbital properties}}

\begin{figure*}
    \centering
    \includegraphics[width=\linewidth]{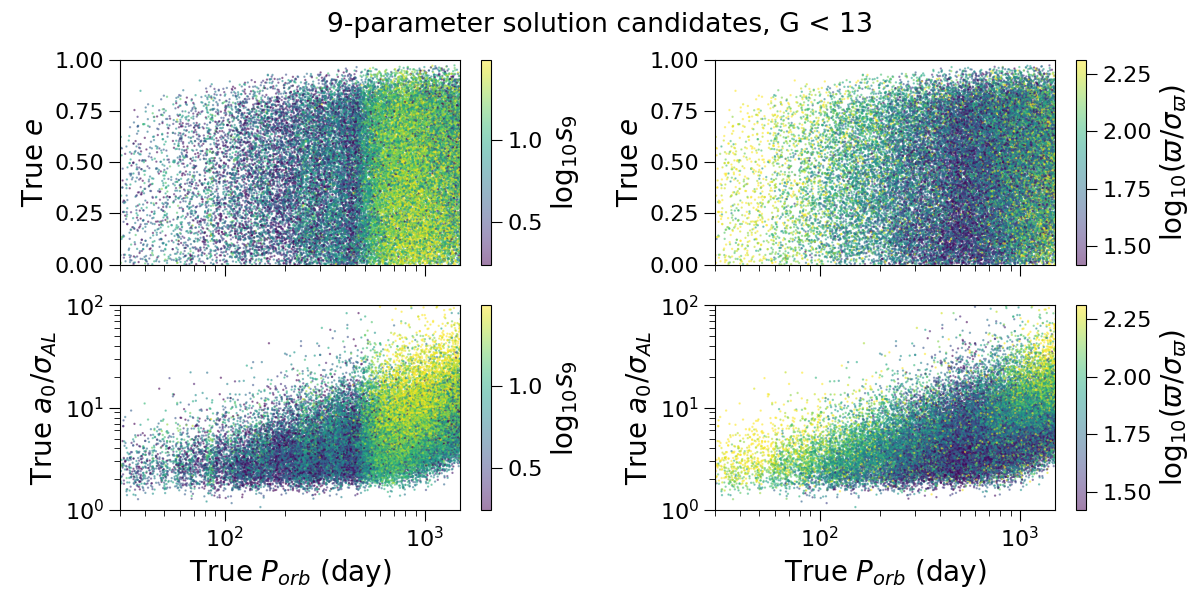}
    \includegraphics[width=\linewidth]{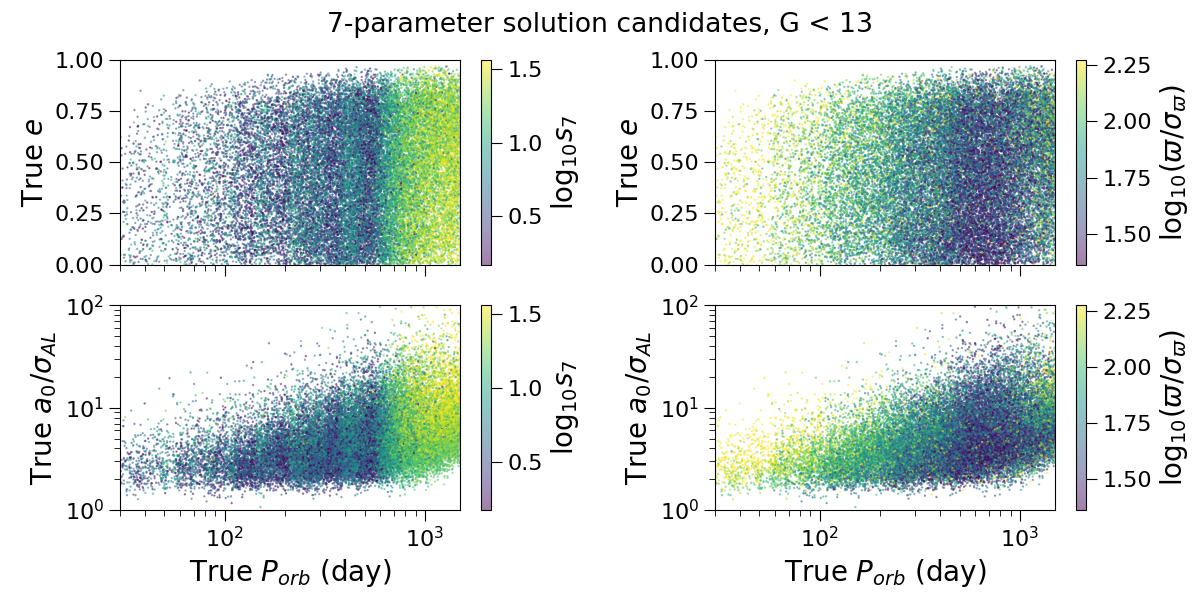}
    \caption{Top four panels: Significance of the 9-parameter acceleration solution (left column) and parallax over error (right column), as a function of true orbital period vs. eccentricity (top row), and orbital period vs. photocenter orbit semimajor axis (bottom row), for fixed noise (i.e., magnitude and visibility periods).
    Bottom four panels: Same as the top four panels, but for the 7-parameter solutions.
    For both the 9-parameter and 7-parameter acceleration solutions, the significance and parallax over error are strong functions of the orbital period and the size of the photocenter orbit semimajor axis, but do not depend on the orbital eccentricity.
    \label{fig:acceleration_significance_parallax_over_error}
    }
\end{figure*}

Here we show the correlations between the criteria for accepting an acceleration model and their dependence on the true orbital properties.
Figure \ref{fig:acceleration_significance_parallax_over_error} demonstrates the relationship between the significances of the 7- and 9-parameter acceleration solutions $s_7, s_9$ and the parallax over errors $\varpi/\sigma_\varpi$ of the respective models.
As described in \S 2.3.2 and \S 4.3 of \cite{Halbwachs:2023}, a 9-parameter solution must have $\varpi/\sigma_\varpi > 2.1 s_9^{1.05}$, $F_2 < 25$, and $s_9 > 12$ to be provisionally accepted in the main processing; a 7-parameter solution must have $\varpi/\sigma_\varpi >1.2 s_7^{1.05}$, $F_2 < 25$, and $s_7 > 12$.
Note that $\varpi/\sigma_\varpi$ is different for each different model (7-parameter acceleration, 9-parameter acceleration, orbital model), as different values are inferred when being fit with different NSS models.

\begin{figure*}
    \centering
    \includegraphics[width=0.49\linewidth]{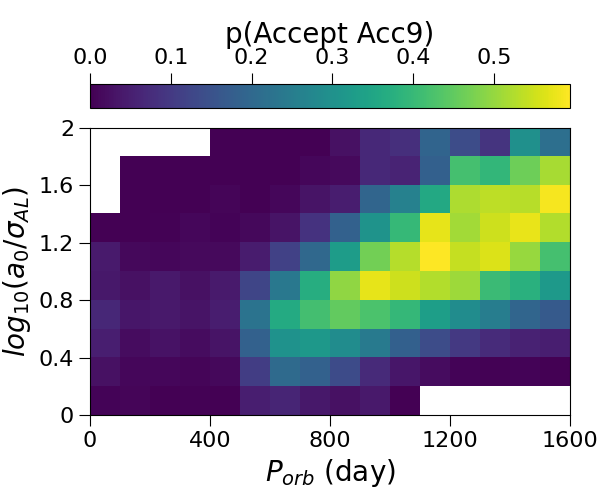}
    \includegraphics[width=0.49\linewidth]{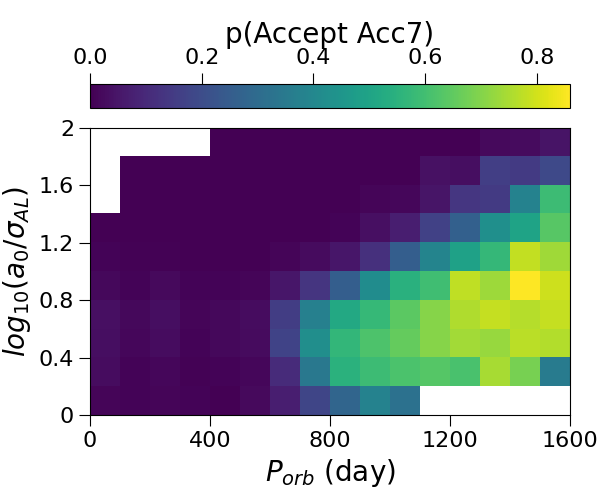}
    \caption{Probability of acceptance of the 9-parameter (left) and 7-parameter acceleration solutions (right), as a function of the true orbital period and the photocenter orbit semimajor axis, scaled by the per-observation uncertainty $\sigma_{AL}$ (which is a function of magnitude).
    The probability of 9-parameter solution acceptance is relative to all sources that have $\ruwe > 1.4$; the probability of 7-parameter solution acceptance is relative to all sources that did not have a 9-parameter solution accepted.}
    \label{fig:acceleration_acceptance}
\end{figure*}

Figure \ref{fig:acceleration_acceptance} shows the probability of a source to be accepted into the 9-parameter and 7-parameter acceleration model branch of the cascade, as a function of the orbital period and photocenter orbit semimajor axis size.
The left panel of Figure \ref{fig:acceleration_acceptance} shows the probability that a source with $\ruwe > 1.4$ is a 9-parameter acceleration solution candidate as a function of $P_{orb}$ and $a_0/\sigma_{AL}$.
The right panel of Figure \ref{fig:acceleration_acceptance} shows the probability that a source that has $\ruwe > 1.4$ and that is not a candidate for a 9-parameter acceleration solution is a candidate for a 7-parameter acceleration solution.
In both cases, longer period solutions $P_{orb} \gtrsim 500$ days are more likely to have an accepted acceleration solution.
For 9-parameter acceleration solutions, intermediate $a_0/\sigma_{AL}$ sources are preferentially accepted, while for the 7-parameter solutions, lower $a_0/\sigma_{AL}$ are preferentially accepted.

\section{Cross-match to external catalogs
\label{app:cross-match to sb9}}

To test the hypothesis that the decrease in solutions with orbital periods $P_{orb} > 550$ days is due to the model cascade and losing solutions to acceleration solutions, we cross-matched the \emph{Gaia} DR3 NSS astrometric orbits and acceleration catalogs to two other binary star catalogs.

\subsection{Cross-match to SB9}
First, we cross-matched the SB9 catalog \citep{Pourbaix:2004} to the \emph{Gaia} DR3 \nssacc\, and \nsstbo\, catalogs to empirically determine the number of orbital solutions lost to the acceleration catalog.\footnote{In \S 6 ``Non-single stars" of \cite{Babusiaux:2023}, they perform the same exercise and find a ``significant fraction" of astrometric acceleration solution could have been fit with an orbital solution, and that the acceleration values disagree from the known values; however, they do not specify how many or what fraction.}

\begin{figure}[h!]
    \centering
    \includegraphics[width=\linewidth]{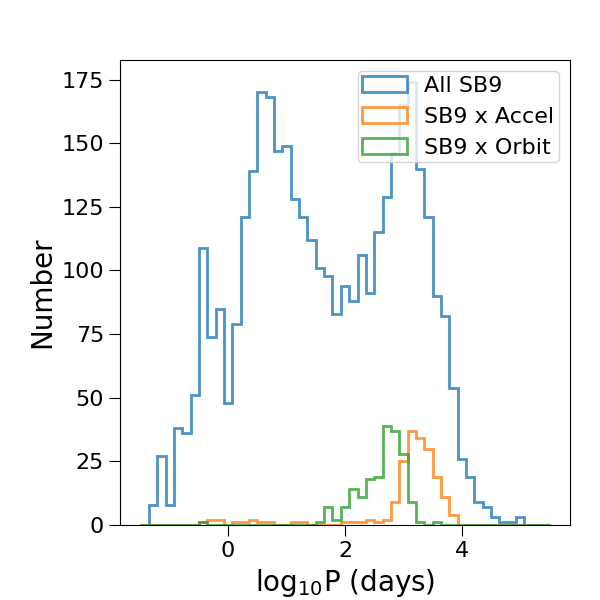}
    \includegraphics[width=\linewidth]{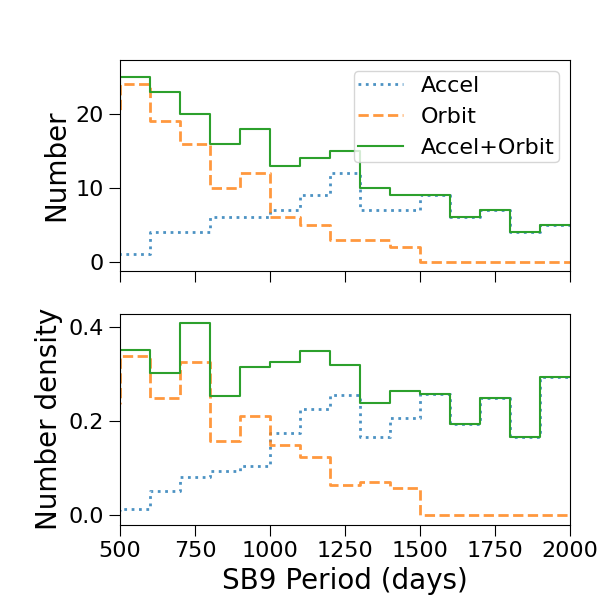}
    \caption{
    \emph{Top panel}: period distribution of SB9 orbits (blue), SB9 orbits cross-matched to the NSS astrometric orbit solutions (green), and SB9 orbits cross-matched to the NSS astrometric acceleration solutions (orange).
    \emph{Middle panel}: Number of NSS astrometric acceleration solutions (blue dot), NSS astrometric orbit solutions (orange dash), and their sum (green solid).
    \emph{Bottom panel}: Same as middle panel, but normalized by the number of total SB9 solutions.
    \label{fig:sb9_acc_tbo}
    }
\end{figure}

The top panel of Figure \ref{fig:sb9_acc_tbo} shows the distribution of orbital periods from SB9, as well as the periods of astrometric acceleration and orbital solutions cross-matched to SB9. 
The middle panel is a histogram of the number of the cross-matched acceleration and orbital solutions in the SB9 catalog with $500 < P_{orb} < 2000$ days.
The number of cross-matched acceleration solutions increases with increasing orbital period, while the number of cross-matched orbital solutions decreases; however, the number of cross-matched acceleration solutions increases slower than the number of cross-matched orbital solutions decreases, as their sum is not flat (middle panel of Figure \ref{fig:sb9_acc_tbo}). 

However, the SB9 catalog is a highly heterogeneous catalog and the period distribution is not representative of an unbiased sample.
To try and mitigate this effect, we also consider the number density of each type of solution; that is, the number of cross-matched orbital or cross-matched acceleration solutions at a given period, divided by the total number of SB9 orbital solutions at that period.
The number density of cross-matched solutions is roughly flat (bottom panel of Figure \ref{fig:sb9_acc_tbo}).

We re-emphasize that the SB9 catalog is a heterogeneous and biased sample, so detailed conclusions cannot be drawn from this test.
However, this test suggests that our hypothesis that the decrease in solutions with $550 < P_{orb} < 1000$ days is due to the model cascade is plausible.

\subsection{Cross-match to WDS-ORB6
\label{app:cross-match to orb6}}

Next, we perform the same cross-matching exercise instead using the Sixth Catalog of Orbits of Visual Binary Stars (Orb6, an updated version of \cite{Hartkopf:2001}).
We used the cross-match of Orb6 to Gaia EDR3 from \cite{Chulkov:2022}, then matched their \emph{Gaia} EDR3 source IDs to the NSS catalogs.
Orb6 binaries are graded based on the reliability of the solution as well as the detection method. 
We only include binaries that have grades 1 (``definitive") or 2 (``good"), or are detected astrometrically or interferometrically. 

The same notes of caution for the SB9 catalog also applies here, although the results are even more tentative here, as the Orb6 orbits are substantially less reliable than the SB9 orbits, even after applying our quality cuts based on cross-match of orbital solutions to available Gaia astrometric orbits.

\begin{figure}[h!]
    \centering
    \includegraphics[width=\linewidth]{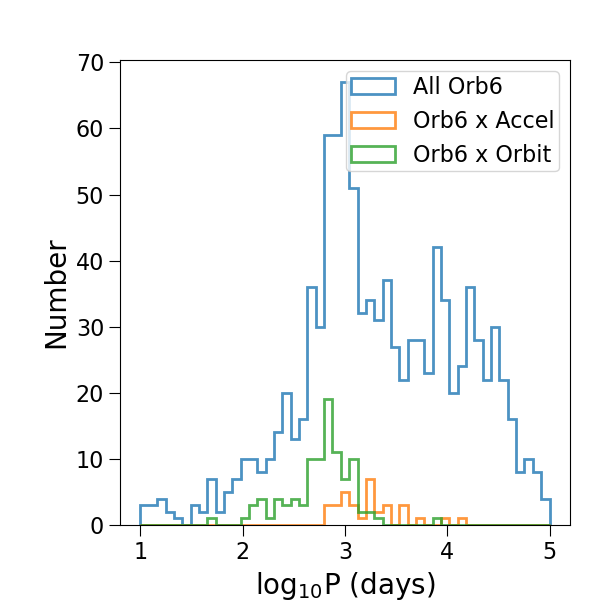}
    \includegraphics[width=\linewidth]{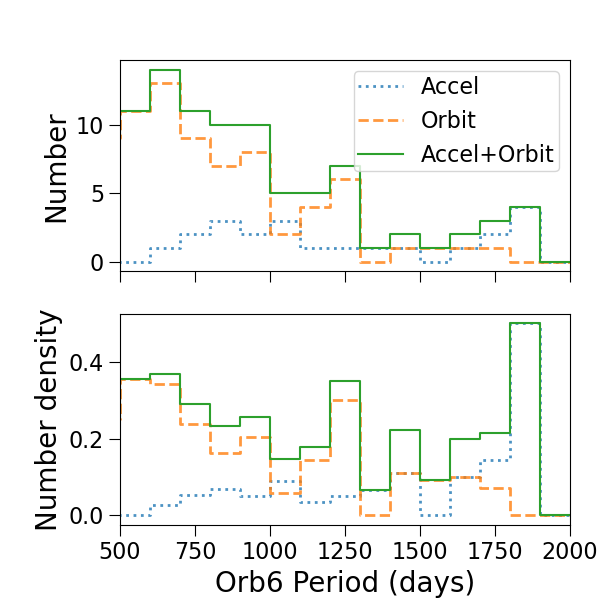}
    \caption{Same as Figure \ref{fig:sb9_acc_tbo}, but for the Orb6 catalog instead of the SB9 catalog.
    \label{fig:orb6_acc_tbo}
    }
\end{figure}

\vspace{-0.6cm}
\bibliography{sample631}{}

\begin{thebibliography}{}
\expandafter\ifx\csname natexlab\endcsname\relax\def\natexlab#1{#1}\fi
\providecommand{\url}[1]{\href{#1}{#1}}
\providecommand{\dodoi}[1]{doi:~\href{http://doi.org/#1}{\nolinkurl{#1}}}
\providecommand{\doeprint}[1]{\href{http://ascl.net/#1}{\nolinkurl{http://ascl.net/#1}}}
\providecommand{\doarXiv}[1]{\href{https://arxiv.org/abs/#1}{\nolinkurl{https://arxiv.org/abs/#1}}}

\bibitem[{{Abbott} {et~al.}(2021){Abbott}, {Abbott}, {Abraham}, {Acernese},
  {Ackley}, {Adams}, {Adams}, {Adhikari}, {Adya}, {Affeldt}, \&
  et~al.}]{Abbott:2021}
{Abbott}, R., {Abbott}, T.~D., {Abraham}, S., {et~al.} 2021, \apjl, 913, L7,
  \dodoi{10.3847/2041-8213/abe949}

\bibitem[{{Andrews} {et~al.}(2022){Andrews}, {Taggart}, \&
  {Foley}}]{Andrews:2022}
{Andrews}, J.~J., {Taggart}, K., \& {Foley}, R. 2022, arXiv e-prints,
  arXiv:2207.00680, \dodoi{10.48550/arXiv.2207.00680}

\bibitem[{{Astropy Collaboration} {et~al.}(2013){Astropy Collaboration},
  {Robitaille}, {Tollerud}, {Greenfield}, {Droettboom}, {Bray}, {Aldcroft},
  {Davis}, {Ginsburg}, {Price-Whelan}, {Kerzendorf}, {Conley}, {Crighton},
  {Barbary}, {Muna}, {Ferguson}, {Grollier}, {Parikh}, {Nair}, {Unther},
  {Deil}, {Woillez}, {Conseil}, {Kramer}, {Turner}, {Singer}, {Fox}, {Weaver},
  {Zabalza}, {Edwards}, {Azalee Bostroem}, {Burke}, {Casey}, {Crawford},
  {Dencheva}, {Ely}, {Jenness}, {Labrie}, {Lim}, {Pierfederici}, {Pontzen},
  {Ptak}, {Refsdal}, {Servillat}, \& {Streicher}}]{Astropy:2013}
{Astropy Collaboration}, {Robitaille}, T.~P., {Tollerud}, E.~J., {et~al.} 2013,
  \aap, 558, A33, \dodoi{10.1051/0004-6361/201322068}

\bibitem[{{Astropy Collaboration} {et~al.}(2018){Astropy Collaboration},
  {Price-Whelan}, {Sip{\H{o}}cz}, {G{\"u}nther}, {Lim}, {Crawford}, {Conseil},
  {Shupe}, {Craig}, {Dencheva}, {Ginsburg}, {Vand erPlas}, {Bradley},
  {P{\'e}rez-Su{\'a}rez}, {de Val-Borro}, {Aldcroft}, {Cruz}, {Robitaille},
  {Tollerud}, {Ardelean}, {Babej}, {Bach}, {Bachetti}, {Bakanov}, {Bamford},
  {Barentsen}, {Barmby}, {Baumbach}, {Berry}, {Biscani}, {Boquien}, {Bostroem},
  {Bouma}, {Brammer}, {Bray}, {Breytenbach}, {Buddelmeijer}, {Burke},
  {Calderone}, {Cano Rodr{\'\i}guez}, {Cara}, {Cardoso}, {Cheedella}, {Copin},
  {Corrales}, {Crichton}, {D'Avella}, {Deil}, {Depagne}, {Dietrich}, {Donath},
  {Droettboom}, {Earl}, {Erben}, {Fabbro}, {Ferreira}, {Finethy}, {Fox},
  {Garrison}, {Gibbons}, {Goldstein}, {Gommers}, {Greco}, {Greenfield},
  {Groener}, {Grollier}, {Hagen}, {Hirst}, {Homeier}, {Horton}, {Hosseinzadeh},
  {Hu}, {Hunkeler}, {Ivezi{\'c}}, {Jain}, {Jenness}, {Kanarek}, {Kendrew},
  {Kern}, {Kerzendorf}, {Khvalko}, {King}, {Kirkby}, {Kulkarni}, {Kumar},
  {Lee}, {Lenz}, {Littlefair}, {Ma}, {Macleod}, {Mastropietro}, {McCully},
  {Montagnac}, {Morris}, {Mueller}, {Mumford}, {Muna}, {Murphy}, {Nelson},
  {Nguyen}, {Ninan}, {N{\"o}the}, {Ogaz}, {Oh}, {Parejko}, {Parley}, {Pascual},
  {Patil}, {Patil}, {Plunkett}, {Prochaska}, {Rastogi}, {Reddy Janga},
  {Sabater}, {Sakurikar}, {Seifert}, {Sherbert}, {Sherwood-Taylor}, {Shih},
  {Sick}, {Silbiger}, {Singanamalla}, {Singer}, {Sladen}, {Sooley},
  {Sornarajah}, {Streicher}, {Teuben}, {Thomas}, {Tremblay}, {Turner},
  {Terr{\'o}n}, {van Kerkwijk}, {de la Vega}, {Watkins}, {Weaver}, {Whitmore},
  {Woillez}, {Zabalza}, \& {Astropy Contributors}}]{Astropy:2018}
{Astropy Collaboration}, {Price-Whelan}, A.~M., {Sip{\H{o}}cz}, B.~M., {et~al.}
  2018, \aj, 156, 123, \dodoi{10.3847/1538-3881/aabc4f}

\bibitem[{{Astropy Collaboration} {et~al.}(2022){Astropy Collaboration},
  {Price-Whelan}, {Lim}, {Earl}, {Starkman}, {Bradley}, {Shupe}, {Patil},
  {Corrales}, {Brasseur}, {N{"o}the}, {Donath}, {Tollerud}, {Morris},
  {Ginsburg}, {Vaher}, {Weaver}, {Tocknell}, {Jamieson}, {van Kerkwijk},
  {Robitaille}, {Merry}, {Bachetti}, {G{"u}nther}, {Aldcroft},
  {Alvarado-Montes}, {Archibald}, {B{'o}di}, {Bapat}, {Barentsen}, {Baz{'a}n},
  {Biswas}, {Boquien}, {Burke}, {Cara}, {Cara}, {Conroy}, {Conseil}, {Craig},
  {Cross}, {Cruz}, {D'Eugenio}, {Dencheva}, {Devillepoix}, {Dietrich},
  {Eigenbrot}, {Erben}, {Ferreira}, {Foreman-Mackey}, {Fox}, {Freij}, {Garg},
  {Geda}, {Glattly}, {Gondhalekar}, {Gordon}, {Grant}, {Greenfield}, {Groener},
  {Guest}, {Gurovich}, {Handberg}, {Hart}, {Hatfield-Dodds}, {Homeier},
  {Hosseinzadeh}, {Jenness}, {Jones}, {Joseph}, {Kalmbach}, {Karamehmetoglu},
  {Ka{l}uszy{'n}ski}, {Kelley}, {Kern}, {Kerzendorf}, {Koch}, {Kulumani},
  {Lee}, {Ly}, {Ma}, {MacBride}, {Maljaars}, {Muna}, {Murphy}, {Norman},
  {O'Steen}, {Oman}, {Pacifici}, {Pascual}, {Pascual-Granado}, {Patil},
  {Perren}, {Pickering}, {Rastogi}, {Roulston}, {Ryan}, {Rykoff}, {Sabater},
  {Sakurikar}, {Salgado}, {Sanghi}, {Saunders}, {Savchenko}, {Schwardt},
  {Seifert-Eckert}, {Shih}, {Jain}, {Shukla}, {Sick}, {Simpson},
  {Singanamalla}, {Singer}, {Singhal}, {Sinha}, {Sip{H{o}}cz}, {Spitler},
  {Stansby}, {Streicher}, {{{S}}umak}, {Swinbank}, {Taranu}, {Tewary},
  {Tremblay}, {Val-Borro}, {Van Kooten}, {Vasovi{'c}}, {Verma}, {de Miranda
  Cardoso}, {Williams}, {Wilson}, {Winkel}, {Wood-Vasey}, {Xue}, {Yoachim},
  {Zhang}, {Zonca}, \& {Astropy Project Contributors}}]{Astropy:2022}
{Astropy Collaboration}, {Price-Whelan}, A.~M., {Lim}, P.~L., {et~al.} 2022,
  \apj, 935, 167, \dodoi{10.3847/1538-4357/ac7c74}

\bibitem[{{Babusiaux} {et~al.}(2023){Babusiaux}, {Fabricius}, {Khanna},
  {Muraveva}, {Reyl{\'e}}, {Spoto}, {Vallenari}, {Luri}, {Arenou},
  {{\'A}lvarez}, {Anders}, {Antoja}, {Balbinot}, {Barache}, {Bauchet},
  {Bossini}, {Busonero}, {Cantat-Gaudin}, {Carrasco}, {Dafonte}, {Diakit{\'e}},
  {Figueras}, {Garcia-Gutierrez}, {Garofalo}, {Helmi}, {Jim{\'e}nez-Arranz},
  {Jordi}, {Kervella}, {Kostrzewa-Rutkowska}, {Leclerc}, {Licata}, {Manteiga},
  {Masip}, {Mongui{\'o}}, {Ramos}, {Robichon}, {Robin}, {Romero-G{\'o}mez},
  {S{\'a}ez}, {Santove{\~n}a}, {Spina}, {Torralba Elipe}, \&
  {Weiler}}]{Babusiaux:2023}
{Babusiaux}, C., {Fabricius}, C., {Khanna}, S., {et~al.} 2023, \aap, 674, A32,
  \dodoi{10.1051/0004-6361/202243790}

\bibitem[{{Belokurov} {et~al.}(2020){Belokurov}, {Penoyre}, {Oh}, {Iorio},
  {Hodgkin}, {Evans}, {Everall}, {Koposov}, {Tout}, {Izzard}, {Clarke}, \&
  {Brown}}]{Belokurov:2020}
{Belokurov}, V., {Penoyre}, Z., {Oh}, S., {et~al.} 2020, \mnras, 496, 1922,
  \dodoi{10.1093/mnras/staa1522}

\bibitem[{{Bovy} {et~al.}(2019){Bovy}, {Leung}, {Hunt}, {Mackereth},
  {Garcia-Hernandez}, \& {Roman-Lopes}}]{Bovy:2019}
{Bovy}, J., {Leung}, H.~W., {Hunt}, J. A.~S., {et~al.} 2019, arXiv e-prints,
  arXiv:1905.11404.
\newblock \doarXiv{1905.11404}

\bibitem[{{Breivik} {et~al.}(2020){Breivik}, {Coughlin}, {Zevin}, {Rodriguez},
  {Kremer}, {Ye}, {Andrews}, {Kurkowski}, {Digman}, {Larson}, \&
  {Rasio}}]{Breivik:2020}
{Breivik}, K., {Coughlin}, S., {Zevin}, M., {et~al.} 2020, \apj, 898, 71,
  \dodoi{10.3847/1538-4357/ab9d85}

\bibitem[{{Chakrabarti} {et~al.}(2023){Chakrabarti}, {Simon}, {Craig},
  {Reggiani}, {Brandt}, {Guhathakurta}, {Dalba}, {Kirby}, {Chang}, {Hey},
  {Savino}, {Geha}, \& {Thompson}}]{Chakrabarti:2023}
{Chakrabarti}, S., {Simon}, J.~D., {Craig}, P.~A., {et~al.} 2023, \aj, 166, 6,
  \dodoi{10.3847/1538-3881/accf21}

\bibitem[{{Chawla} {et~al.}(2022){Chawla}, {Chatterjee}, {Breivik}, {Moorthy},
  {Andrews}, \& {Sanderson}}]{Chawla:2022}
{Chawla}, C., {Chatterjee}, S., {Breivik}, K., {et~al.} 2022, \apj, 931, 107,
  \dodoi{10.3847/1538-4357/ac60a5}

\bibitem[{{Choi} {et~al.}(2016){Choi}, {Dotter}, {Conroy}, {Cantiello},
  {Paxton}, \& {Johnson}}]{Choi:2016}
{Choi}, J., {Dotter}, A., {Conroy}, C., {et~al.} 2016, \apj, 823, 102,
  \dodoi{10.3847/0004-637X/823/2/102}

\bibitem[{{Chulkov} \& {Malkov}(2022)}]{Chulkov:2022}
{Chulkov}, D., \& {Malkov}, O. 2022, \mnras, 517, 2925,
  \dodoi{10.1093/mnras/stac2827}

\bibitem[{{Dotter}(2016)}]{Dotter:2016}
{Dotter}, A. 2016, \apjs, 222, 8, \dodoi{10.3847/0067-0049/222/1/8}

\bibitem[{{Drimmel} {et~al.}(2003){Drimmel}, {Cabrera-Lavers}, \&
  {L{\'o}pez-Corredoira}}]{Drimmel:2003}
{Drimmel}, R., {Cabrera-Lavers}, A., \& {L{\'o}pez-Corredoira}, M. 2003, \aap,
  409, 205, \dodoi{10.1051/0004-6361:20031070}

\bibitem[{{Duquennoy} \& {Mayor}(1991)}]{Duquennoy:1991}
{Duquennoy}, A., \& {Mayor}, M. 1991, \aap, 248, 485

\bibitem[{{El-Badry}(2024)}]{El-Badry:2024}
{El-Badry}, K. 2024, \nar, 98, 101694, \dodoi{10.1016/j.newar.2024.101694}

\bibitem[{{El-Badry} {et~al.}(2024){El-Badry}, {Lam}, {Holl}, {Halbwachs},
  {Rix}, {Mazeh}, \& {Shahaf}}]{Kareem}
{El-Badry}, K., {Lam}, C., {Holl}, B., {et~al.} 2024, The Open Journal of
  Astrophysics, 7, 100, \dodoi{10.33232/001c.125461}

\bibitem[{{El-Badry} {et~al.}(2023{\natexlab{a}}){El-Badry}, {Rix}, {Quataert},
  {Howard}, {Isaacson}, {Fuller}, {Hawkins}, {Breivik}, {Wong}, {Rodriguez},
  {Conroy}, {Shahaf}, {Mazeh}, {Arenou}, {Burdge}, {Bashi}, {Faigler}, {Weisz},
  {Seeburger}, {Almada Monter}, \& {Wojno}}]{El-Badry:2023a}
{El-Badry}, K., {Rix}, H.-W., {Quataert}, E., {et~al.} 2023{\natexlab{a}},
  \mnras, 518, 1057, \dodoi{10.1093/mnras/stac3140}

\bibitem[{{El-Badry} {et~al.}(2023{\natexlab{b}}){El-Badry}, {Rix}, {Cendes},
  {Rodriguez}, {Conroy}, {Quataert}, {Hawkins}, {Zari}, {Hobson}, {Breivik},
  {Rau}, {Berger}, {Shahaf}, {Seeburger}, {Burdge}, {Latham}, {Buchhave},
  {Bieryla}, {Bashi}, {Mazeh}, \& {Faigler}}]{El-Badry:2023b}
{El-Badry}, K., {Rix}, H.-W., {Cendes}, Y., {et~al.} 2023{\natexlab{b}},
  \mnras, 521, 4323, \dodoi{10.1093/mnras/stad799}

\bibitem[{{Gaia Collaboration} {et~al.}(2016{\natexlab{a}}){Gaia
  Collaboration}, {Prusti}, {de Bruijne}, {Brown}, {Vallenari}, {Babusiaux},
  {Bailer-Jones}, {Bastian}, {Biermann}, {Evans}, {Eyer}, {Jansen}, {Jordi},
  {Klioner}, {Lammers}, {Lindegren}, {Luri}, {Mignard}, {Milligan}, {Panem},
  {Poinsignon}, {Pourbaix}, {Randich}, {Sarri}, {Sartoretti}, {Siddiqui},
  {Soubiran}, {Valette}, {van Leeuwen}, {Walton}, {Aerts}, {Arenou}, {Cropper},
  {Drimmel}, {H{\o}g}, {Katz}, {Lattanzi}, {O'Mullane}, {Grebel}, {Holland},
  {Huc}, {Passot}, {Bramante}, {Cacciari}, {Casta{\~n}eda}, {Chaoul}, {Cheek},
  {De Angeli}, {Fabricius}, {Guerra}, {Hern{\'a}ndez}, {Jean-Antoine-Piccolo},
  {Masana}, {Messineo}, {Mowlavi}, {Nienartowicz}, {Ord{\'o}{\~n}ez-Blanco},
  {Panuzzo}, {Portell}, {Richards}, {Riello}, {Seabroke}, {Tanga},
  {Th{\'e}venin}, {Torra}, {Els}, {Gracia-Abril}, {Comoretto},
  {Garcia-Reinaldos}, {Lock}, \& {Mercier}}]{Prusti:2016}
{Gaia Collaboration}, {Prusti}, T., {de Bruijne}, J.~H.~J., {et~al.}
  2016{\natexlab{a}}, \aap, 595, A1, \dodoi{10.1051/0004-6361/201629272}

\bibitem[{{Gaia Collaboration} {et~al.}(2016{\natexlab{b}}){Gaia
  Collaboration}, {Brown}, {Vallenari}, {Prusti}, {de Bruijne}, {Mignard},
  {Drimmel}, {Babusiaux}, {Bailer-Jones}, {Bastian}, {Biermann}, {Evans},
  {Eyer}, {Jansen}, {Jordi}, {Katz}, {Klioner}, {Lammers}, {Lindegren}, {Luri},
  {O'Mullane}, {Panem}, {Pourbaix}, {Randich}, {Sartoretti}, {Siddiqui},
  {Soubiran}, {Valette}, {van Leeuwen}, {Walton}, {Aerts}, {Arenou}, {Cropper},
  {H{\o}g}, {Lattanzi}, {Grebel}, {Holland}, {Huc}, {Passot}, {Perryman},
  {Bramante}, {Cacciari}, {Casta{\~n}eda}, {Chaoul}, {Cheek}, {De Angeli},
  {Fabricius}, {Guerra}, {Hern{\'a}ndez}, {Jean-Antoine-Piccolo}, {Masana},
  {Messineo}, {Mowlavi}, {Nienartowicz}, {Ord{\'o}{\~n}ez-Blanco}, {Panuzzo},
  {Portell}, {Richards}, {Riello}, {Seabroke}, {Tanga}, {Th{\'e}venin},
  {Torra}, {Els}, {Gracia-Abril}, {Comoretto}, {Garcia-Reinaldos}, {Lock}, \&
  {Mercier}}]{GaiaDR1:2016}
{Gaia Collaboration}, {Brown}, A.~G.~A., {Vallenari}, A., {et~al.}
  2016{\natexlab{b}}, \aap, 595, A2, \dodoi{10.1051/0004-6361/201629512}

\bibitem[{{Gaia Collaboration} {et~al.}(2018){Gaia Collaboration}, {Brown},
  {Vallenari}, {Prusti}, {de Bruijne}, {Babusiaux}, {Bailer-Jones}, {Biermann},
  {Evans}, {Eyer}, {Jansen}, {Jordi}, {Klioner}, {Lammers}, {Lindegren},
  {Luri}, {Mignard}, {Panem}, {Pourbaix}, {Randich}, {Sartoretti}, {Siddiqui},
  {Soubiran}, {van Leeuwen}, {Walton}, {Arenou}, {Bastian}, {Cropper},
  {Drimmel}, {Katz}, {Lattanzi}, {Bakker}, {Cacciari}, {Casta{\~n}eda},
  {Chaoul}, {Cheek}, {De Angeli}, {Fabricius}, {Guerra}, {Holl}, {Masana},
  {Messineo}, {Mowlavi}, {Nienartowicz}, {Panuzzo}, {Portell}, {Riello},
  {Seabroke}, {Tanga}, {Th{\'e}venin}, {Gracia-Abril}, {Comoretto}, \&
  {Garcia-Reinaldos}}]{GaiaDR2:2018}
---. 2018, \aap, 616, A1, \dodoi{10.1051/0004-6361/201833051}

\bibitem[{{Gaia Collaboration} {et~al.}(2021{\natexlab{a}}){Gaia
  Collaboration}, {Brown}, {Vallenari}, {Prusti}, {de Bruijne}, {Babusiaux},
  {Biermann}, {Creevey}, {Evans}, {Eyer}, {Hutton}, {Jansen}, {Jordi},
  {Klioner}, {Lammers}, {Lindegren}, {Luri}, {Mignard}, {Panem}, {Pourbaix},
  {Randich}, {Sartoretti}, {Soubiran}, {Walton}, {Arenou}, {Bailer-Jones},
  {Bastian}, {Cropper}, {Drimmel}, {Katz}, {Lattanzi}, {van Leeuwen}, {Bakker},
  {Cacciari}, {Casta{\~n}eda}, {De Angeli}, {Ducourant}, {Fabricius},
  {Fouesneau}, {Fr{\'e}mat}, {Guerra}, {Guerrier}, {Guiraud}, {Jean-Antoine
  Piccolo}, {Masana}, {Messineo}, {Mowlavi}, {Nicolas}, {Nienartowicz},
  {Pailler}, {Panuzzo}, {Riclet}, {Roux}, {Seabroke}, {Sordo}, {Tanga},
  {Th{\'e}venin}, {Gracia-Abril}, {Portell}, \& {Teyssier}}]{GaiaEDR3:2021}
---. 2021{\natexlab{a}}, \aap, 649, A1, \dodoi{10.1051/0004-6361/202039657}

\bibitem[{{Gaia Collaboration} {et~al.}(2021{\natexlab{b}}){Gaia
  Collaboration}, {Smart}, {Sarro}, {Rybizki}, {Reyl{\'e}}, {Robin}, {Hambly},
  {Abbas}, {Barstow}, {de Bruijne}, {Bucciarelli}, {Carrasco}, {Cooper},
  {Hodgkin}, {Masana}, {Michalik}, {Sahlmann}, {Sozzetti}, {Brown},
  {Vallenari}, {Prusti}, {Babusiaux}, {Biermann}, {Creevey}, {Evans}, {Eyer},
  {Hutton}, {Jansen}, {Jordi}, {Klioner}, {Lammers}, {Lindegren}, {Luri},
  {Mignard}, {Panem}, {Pourbaix}, {Randich}, {Sartoretti}, {Soubiran}, \&
  {Walton}}]{Smart:2021}
{Gaia Collaboration}, {Smart}, R.~L., {Sarro}, L.~M., {et~al.}
  2021{\natexlab{b}}, \aap, 649, A6, \dodoi{10.1051/0004-6361/202039498}

\bibitem[{{Gaia Collaboration} {et~al.}(2023{\natexlab{a}}){Gaia
  Collaboration}, {Vallenari}, {Brown}, {Prusti}, {de Bruijne}, {Arenou},
  {Babusiaux}, {Biermann}, {Creevey}, {Ducourant}, {Evans}, {Eyer}, {Guerra},
  {Hutton}, {Jordi}, {Klioner}, {Lammers}, {Lindegren}, {Luri}, {Mignard},
  {Panem}, {Pourbaix}, {Randich}, {Sartoretti}, {Soubiran}, {Tanga}, {Walton},
  {Bailer-Jones}, {Bastian}, {Drimmel}, {Jansen}, {Katz}, {Lattanzi}, {van
  Leeuwen}, {Bakker}, {Cacciari}, {Casta{\~n}eda}, {De Angeli}, {Fabricius},
  {Fouesneau}, {Fr{\'e}mat}, {Galluccio}, {Guerrier}, {Heiter}, {Masana},
  {Messineo}, {Mowlavi}, {Nicolas}, {Nienartowicz}, {Pailler}, {Panuzzo},
  {Riclet}, {Roux}, {Seabroke}, {Sordo}, {Th{\'e}venin}, {Gracia-Abril},
  {Portell}, \& {Teyssier}}]{GaiaDR3:2023}
{Gaia Collaboration}, {Vallenari}, A., {Brown}, A.~G.~A., {et~al.}
  2023{\natexlab{a}}, \aap, 674, A1, \dodoi{10.1051/0004-6361/202243940}

\bibitem[{{Gaia Collaboration} {et~al.}(2023{\natexlab{b}}){Gaia
  Collaboration}, {Arenou}, {Babusiaux}, {Barstow}, {Faigler}, {Jorissen},
  {Kervella}, {Mazeh}, {Mowlavi}, {Panuzzo}, {Sahlmann}, {Shahaf}, {Sozzetti},
  {Bauchet}, {Damerdji}, {Gavras}, {Giacobbe}, {Gosset}, {Halbwachs}, {Holl},
  {Lattanzi}, {Leclerc}, {Morel}, {Pourbaix}, {Re Fiorentin}, {Sadowski},
  {S{\'e}gransan}, {Siopis}, {Teyssier}, {Zwitter}, {Planquart}, \&
  {Brown}}]{Arenou:2023}
{Gaia Collaboration}, {Arenou}, F., {Babusiaux}, C., {et~al.}
  2023{\natexlab{b}}, \aap, 674, A34, \dodoi{10.1051/0004-6361/202243782}

\bibitem[{{Gaia Collaboration} {et~al.}(2024){Gaia Collaboration}, {Panuzzo},
  {Mazeh}, {Arenou}, {Holl}, {Caffau}, {Jorissen}, {Babusiaux}, {Gavras},
  {Sahlmann}, {Bastian}, {Wyrzykowski}, {Eyer}, {Leclerc}, {Bauchet},
  {Bombrun}, {Mowlavi}, {Seabroke}, {Teyssier}, {Balbinot}, {Helmi}, {Brown},
  {Vallenari}, {Prusti}, {de Bruijne}, {Barbier}, {Biermann}, {Creevey},
  {Ducourant}, {Evans}, {Guerra}, {Hutton}, {Jordi}, {Klioner}, {Lammers},
  {Lindegren}, {Luri}, {Mignard}, {Nicolas}, {Randich}, {Sartoretti},
  {Smiljanic}, {Tanga}, {Walton}, {Aerts}, {Bailer-Jones}, {Cropper},
  {Drimmel}, {Jansen}, {Katz}, {Lattanzi}, {Soubiran}, {Th{\'e}venin}, {van
  Leeuwen}, {Andrae}, {Audard}, {Bakker}, {Blomme}, {Casta{\~n}eda}, {De
  Angeli}, {Fabricius}, {Fouesneau}, {Fr{\'e}mat}, {Galluccio}, {Guerrier},
  {Heiter}, {Masana}, {Messineo}, {Nienartowicz}, {Pailler}, {Riclet}, {Roux},
  {Sordo}, {Gracia-Abril}, {Portell}, {Altmann}, {Benson}, {Berthier},
  {Burgess}, {Busonero}, {Busso}, {Cacciari}, {C{\'a}novas}, {Carrasco},
  {Carry}, {Cellino}, {Cheek}, {Clementini}, {Damerdji}, {Davidson}, {de
  Teodoro}, {Delchambre}, {Dell'Oro}, {Fraile Garcia}, {Garabato},
  {Garc{\'\i}a-Lario}, {Haigron}, {Hambly}, {Harrison}, {Hatzidimitriou},
  {Hern{\'a}ndez}, {Hestroffer}, {Hodgkin}, {Jamal}, {Jevardat de Fombelle},
  {Jordan}, {Krone-Martins}, {Lanzafame}, {L{\"o}ffler}, {Lorca}, {Marchal},
  {Marrese}, {Moitinho}, {Muinonen}, {Nu{\~n}ez Campos}, {Oreshina-Slezak},
  {Osborne}, {Pancino}, {Pauwels}, {Recio-Blanco}, {Riello}, {Rimoldini},
  {Robin}, {Roegiers}, {Sarro}, \& {Schultheis}}]{Panuzzo:2024}
{Gaia Collaboration}, {Panuzzo}, P., {Mazeh}, T., {et~al.} 2024, \aap, 686, L2,
  \dodoi{10.1051/0004-6361/202449763}

\bibitem[{{Gosset} {et~al.}(2024){Gosset}, {Damerdji}, {Morel}, {Delchambre},
  {Halbwachs}, {Sadowski}, {Pourbaix}, {Sozzetti}, {Panuzzo}, \&
  {Arenou}}]{Gosset:2024}
{Gosset}, E., {Damerdji}, Y., {Morel}, T., {et~al.} 2024, arXiv e-prints,
  arXiv:2410.14372.
\newblock \doarXiv{2410.14372}

\bibitem[{{Green} {et~al.}(2019){Green}, {Schlafly}, {Zucker}, {Speagle}, \&
  {Finkbeiner}}]{Green:2019}
{Green}, G.~M., {Schlafly}, E., {Zucker}, C., {Speagle}, J.~S., \&
  {Finkbeiner}, D. 2019, \apj, 887, 93, \dodoi{10.3847/1538-4357/ab5362}

\bibitem[{{Halbwachs} {et~al.}(2023){Halbwachs}, {Pourbaix}, {Arenou},
  {Galluccio}, {Guillout}, {Bauchet}, {Marchal}, {Sadowski}, \&
  {Teyssier}}]{Halbwachs:2023}
{Halbwachs}, J.-L., {Pourbaix}, D., {Arenou}, F., {et~al.} 2023, \aap, 674, A9,
  \dodoi{10.1051/0004-6361/202243969}

\bibitem[{{Hartkopf} {et~al.}(2001){Hartkopf}, {Mason}, \&
  {Worley}}]{Hartkopf:2001}
{Hartkopf}, W.~I., {Mason}, B.~D., \& {Worley}, C.~E. 2001, \aj, 122, 3472,
  \dodoi{10.1086/323921}

\bibitem[{{Holl} {et~al.}(2023){Holl}, {Fabricius}, {Portell}, {Lindegren},
  {Panuzzo}, {Bernet}, {Casta{\~n}eda}, {Jevardat de Fombelle}, {Audard},
  {Ducourant}, {Harrison}, {Evans}, {Busso}, {Sozzetti}, {Gosset}, {Arenou},
  {De Angeli}, {Riello}, {Eyer}, {Rimoldini}, {Gavras}, {Mowlavi},
  {Nienartowicz}, {Lecoeur-Ta{\"\i}bi}, {Garc{\'\i}a-Lario}, \&
  {Pourbaix}}]{Holl:2023}
{Holl}, B., {Fabricius}, C., {Portell}, J., {et~al.} 2023, \aap, 674, A25,
  \dodoi{10.1051/0004-6361/202245353}

\bibitem[{{Hunter}(2007)}]{Hunter:2007}
{Hunter}, J.~D. 2007, Computing in Science \& Engineering, 9, 90,
  \dodoi{10.1109/MCSE.2007.55}

\bibitem[{{Kalirai} {et~al.}(2008){Kalirai}, {Hansen}, {Kelson}, {Reitzel},
  {Rich}, \& {Richer}}]{Kalirai:2008}
{Kalirai}, J.~S., {Hansen}, B.~M.~S., {Kelson}, D.~D., {et~al.} 2008, \apj,
  676, 594, \dodoi{10.1086/527028}

\bibitem[{{Kirk} {et~al.}(2016){Kirk}, {Conroy}, {Pr{\v{s}}a}, {Abdul-Masih},
  {Kochoska}, {Matijevi{\v{c}}}, {Hambleton}, {Barclay}, {Bloemen}, {Boyajian},
  {Doyle}, {Fulton}, {Hoekstra}, {Jek}, {Kane}, {Kostov}, {Latham}, {Mazeh},
  {Orosz}, {Pepper}, {Quarles}, {Ragozzine}, {Shporer}, {Southworth},
  {Stassun}, {Thompson}, {Welsh}, {Agol}, {Derekas}, {Devor}, {Fischer},
  {Green}, {Gropp}, {Jacobs}, {Johnston}, {LaCourse}, {Saetre}, {Schwengeler},
  {Toczyski}, {Werner}, {Garrett}, {Gore}, {Martinez}, {Spitzer}, {Stevick},
  {Thomadis}, {Vrijmoet}, {Yenawine}, {Batalha}, \& {Borucki}}]{Kirk:2016}
{Kirk}, B., {Conroy}, K., {Pr{\v{s}}a}, A., {et~al.} 2016, \aj, 151, 68,
  \dodoi{10.3847/0004-6256/151/3/68}

\bibitem[{{Kroupa}(2001)}]{Kroupa:2001}
{Kroupa}, P. 2001, \mnras, 322, 231, \dodoi{10.1046/j.1365-8711.2001.04022.x}

\bibitem[{{Laithwaite} \& {Warren}(2020)}]{Laithwaite:2020}
{Laithwaite}, R.~C., \& {Warren}, S.~J. 2020, \mnras, 499, 2587,
  \dodoi{10.1093/mnras/staa2979}

\bibitem[{Lindegren(2018)}]{Lindegren:2018}
Lindegren, L. 2018.
\newblock \url{http://www.rssd.esa.int/doc_fetch.php?id=3757412}

\bibitem[{{Lindegren} {et~al.}(2012){Lindegren}, {Lammers}, {Hobbs},
  {O'Mullane}, {Bastian}, \& {Hern{\'a}ndez}}]{Lindegren:2012}
{Lindegren}, L., {Lammers}, U., {Hobbs}, D., {et~al.} 2012, \aap, 538, A78,
  \dodoi{10.1051/0004-6361/201117905}

\bibitem[{{Lindegren} {et~al.}(2018){Lindegren}, {Hern{\'a}ndez}, {Bombrun},
  {Klioner}, {Bastian}, {Ramos-Lerate}, {de Torres}, {Steidelm{\"u}ller},
  {Stephenson}, {Hobbs}, {Lammers}, {Biermann}, {Geyer}, {Hilger}, {Michalik},
  {Stampa}, {McMillan}, {Casta{\~n}eda}, {Clotet}, {Comoretto}, {Davidson},
  {Fabricius}, {Gracia}, {Hambly}, {Hutton}, {Mora}, {Portell}, {van Leeuwen},
  {Abbas}, {Abreu}, {Altmann}, {Andrei}, {Anglada}, {Balaguer-N{\'u}{\~n}ez},
  {Barache}, {Becciani}, {Bertone}, {Bianchi}, {Bouquillon}, {Bourda},
  {Br{\"u}semeister}, {Bucciarelli}, {Busonero}, {Buzzi}, {Cancelliere},
  {Carlucci}, {Charlot}, {Cheek}, {Crosta}, {Crowley}, {de Bruijne}, {de
  Felice}, {Drimmel}, {Esquej}, {Fienga}, {Fraile}, {Gai}, {Garralda},
  {Gonz{\'a}lez-Vidal}, {Guerra}, {Hauser}, {Hofmann}, {Holl}, {Jordan},
  {Lattanzi}, {Lenhardt}, {Liao}, {Licata}, {Lister}, {L{\"o}ffler},
  {Marchant}, {Martin-Fleitas}, {Messineo}, {Mignard}, {Morbidelli}, {Poggio},
  {Riva}, {Rowell}, {Salguero}, {Sarasso}, {Sciacca}, {Siddiqui}, {Smart},
  {Spagna}, {Steele}, {Taris}, {Torra}, {van Elteren}, {van Reeven}, \&
  {Vecchiato}}]{Lindegren_etal:2018}
{Lindegren}, L., {Hern{\'a}ndez}, J., {Bombrun}, A., {et~al.} 2018, \aap, 616,
  A2, \dodoi{10.1051/0004-6361/201832727}

\bibitem[{{Lindegren} {et~al.}(2021){Lindegren}, {Klioner}, {Hern{\'a}ndez},
  {Bombrun}, {Ramos-Lerate}, {Steidelm{\"u}ller}, {Bastian}, {Biermann}, {de
  Torres}, {Gerlach}, {Geyer}, {Hilger}, {Hobbs}, {Lammers}, {McMillan},
  {Stephenson}, {Casta{\~n}eda}, {Davidson}, {Fabricius}, {Gracia-Abril},
  {Portell}, {Rowell}, {Teyssier}, {Torra}, {Bartolom{\'e}}, {Clotet},
  {Garralda}, {Gonz{\'a}lez-Vidal}, {Torra}, {Abbas}, {Altmann}, {Anglada
  Varela}, {Balaguer-N{\'u}{\~n}ez}, {Balog}, {Barache}, {Becciani}, {Bernet},
  {Bertone}, {Bianchi}, {Bouquillon}, {Brown}, {Bucciarelli}, {Busonero},
  {Butkevich}, {Buzzi}, {Cancelliere}, {Carlucci}, {Charlot}, {Cioni},
  {Crosta}, {Crowley}, {del Peloso}, {del Pozo}, {Drimmel}, {Esquej}, {Fienga},
  {Fraile}, {Gai}, {Garcia-Reinaldos}, {Guerra}, {Hambly}, {Hauser},
  {Jan{\ss}en}, {Jordan}, {Kostrzewa-Rutkowska}, {Lattanzi}, {Liao}, {Licata},
  {Lister}, {L{\"o}ffler}, {Marchant}, {Masip}, {Mignard}, {Mints}, {Molina},
  {Mora}, {Morbidelli}, {Murphy}, {Pagani}, {Panuzzo}, {Pe{\~n}alosa Esteller},
  {Poggio}, {Re Fiorentin}, {Riva}, {Sagrist{\`a} Sell{\'e}s}, {Sanchez
  Gimenez}, {Sarasso}, {Sciacca}, {Siddiqui}, {Smart}, {Souami}, {Spagna},
  {Steele}, {Taris}, {Utrilla}, {van Reeven}, \& {Vecchiato}}]{Lindegren:2021a}
{Lindegren}, L., {Klioner}, S.~A., {Hern{\'a}ndez}, J., {et~al.} 2021, \aap,
  649, A2, \dodoi{10.1051/0004-6361/202039709}

\bibitem[{{Mar{\'\i}n Pina} {et~al.}(2024){Mar{\'\i}n Pina}, {Rastello},
  {Gieles}, {Kremer}, {Fitzgerald}, \& {Rando Forastier}}]{MarinPina:2024}
{Mar{\'\i}n Pina}, D., {Rastello}, S., {Gieles}, M., {et~al.} 2024, \aap, 688,
  L2, \dodoi{10.1051/0004-6361/202450460}

\bibitem[{{Marshall} {et~al.}(2006){Marshall}, {Robin}, {Reyl{\'e}},
  {Schultheis}, \& {Picaud}}]{Marshall:2006}
{Marshall}, D.~J., {Robin}, A.~C., {Reyl{\'e}}, C., {Schultheis}, M., \&
  {Picaud}, S. 2006, \aap, 453, 635, \dodoi{10.1051/0004-6361:20053842}

\bibitem[{{Moe} \& {Di Stefano}(2015)}]{Moe:2015}
{Moe}, M., \& {Di Stefano}, R. 2015, \apj, 810, 61,
  \dodoi{10.1088/0004-637X/810/1/61}

\bibitem[{{Moe} \& {Di Stefano}(2017)}]{Moe:2017}
---. 2017, \apjs, 230, 15, \dodoi{10.3847/1538-4365/aa6fb6}

\bibitem[{{Morton}(2015)}]{Morton:2015}
{Morton}, T.~D. 2015, {isochrones: Stellar model grid package}, Astrophysics
  Source Code Library, record ascl:1503.010

\bibitem[{{Nagarajan} {et~al.}(2024){Nagarajan}, {El-Badry}, {Triaud},
  {Baycroft}, {Latham}, {Bieryla}, {Buchhave}, {Rix}, {Quataert}, {Howard},
  {Isaacson}, \& {Hobson}}]{Nagarajan:2024}
{Nagarajan}, P., {El-Badry}, K., {Triaud}, A. H.~M.~J., {et~al.} 2024, \pasp,
  136, 014202, \dodoi{10.1088/1538-3873/ad1ba7}

\bibitem[{{Nagarajan} {et~al.}(2025){Nagarajan}, {El-Badry}, {Chawla},
  {Niccol{\`o} Di Carlo}, {Breivik}, {Rodriguez}, {Agrawal}, {Delfavero}, \&
  {Chatterjee}}]{Nagarajan:2025}
{Nagarajan}, P., {El-Badry}, K., {Chawla}, C., {et~al.} 2025, arXiv e-prints,
  arXiv:2502.03527, \dodoi{10.48550/arXiv.2502.03527}

\bibitem[{{Penoyre} {et~al.}(2020){Penoyre}, {Belokurov}, {Wyn Evans},
  {Everall}, \& {Koposov}}]{Penoyre:2020}
{Penoyre}, Z., {Belokurov}, V., {Wyn Evans}, N., {Everall}, A., \& {Koposov},
  S.~E. 2020, \mnras, 495, 321, \dodoi{10.1093/mnras/staa1148}

\bibitem[{{Pourbaix} {et~al.}(2004){Pourbaix}, {Tokovinin}, {Batten}, {Fekel},
  {Hartkopf}, {Levato}, {Morrell}, {Torres}, \& {Udry}}]{Pourbaix:2004}
{Pourbaix}, D., {Tokovinin}, A.~A., {Batten}, A.~H., {et~al.} 2004, \aap, 424,
  727, \dodoi{10.1051/0004-6361:20041213}

\bibitem[{{Raghavan} {et~al.}(2010){Raghavan}, {McAlister}, {Henry}, {Latham},
  {Marcy}, {Mason}, {Gies}, {White}, \& {ten Brummelaar}}]{Raghavan:2010}
{Raghavan}, D., {McAlister}, H.~A., {Henry}, T.~J., {et~al.} 2010, \apjs, 190,
  1, \dodoi{10.1088/0067-0049/190/1/1}

\bibitem[{{Reipurth} \& {Zinnecker}(1993)}]{Reipurth:1993}
{Reipurth}, B., \& {Zinnecker}, H. 1993, \aap, 278, 81

\bibitem[{{Robin} {et~al.}(2003){Robin}, {Reyl{\'e}}, {Derri{\`e}re}, \&
  {Picaud}}]{Robin:2003}
{Robin}, A.~C., {Reyl{\'e}}, C., {Derri{\`e}re}, S., \& {Picaud}, S. 2003,
  \aap, 409, 523, \dodoi{10.1051/0004-6361:20031117}

\bibitem[{{Rowan} {et~al.}(2022){Rowan}, {Jayasinghe}, {Stanek}, {Kochanek},
  {Thompson}, {Shappee}, {Holoien}, {Prieto}, \& {Giles}}]{Rowan:2022}
{Rowan}, D.~M., {Jayasinghe}, T., {Stanek}, K.~Z., {et~al.} 2022, \mnras, 517,
  2190, \dodoi{10.1093/mnras/stac2520}

\bibitem[{{Shahaf} {et~al.}(2023){Shahaf}, {Bashi}, {Mazeh}, {Faigler},
  {Arenou}, {El-Badry}, \& {Rix}}]{Shahaf:2023}
{Shahaf}, S., {Bashi}, D., {Mazeh}, T., {et~al.} 2023, \mnras, 518, 2991,
  \dodoi{10.1093/mnras/stac3290}

\bibitem[{{Sharma} {et~al.}(2011){Sharma}, {Bland-Hawthorn}, {Johnston}, \&
  {Binney}}]{Sharma:2011}
{Sharma}, S., {Bland-Hawthorn}, J., {Johnston}, K.~V., \& {Binney}, J. 2011,
  \apj, 730, 3, \dodoi{10.1088/0004-637X/730/1/3}

\bibitem[{{Soszy{\'n}ski} {et~al.}(2016){Soszy{\'n}ski}, {Pawlak},
  {Pietrukowicz}, {Udalski}, {Szyma{\'n}ski}, {Wyrzykowski}, {Ulaczyk},
  {Poleski}, {Koz{\l}owski}, {Skowron}, {Skowron}, {Mr{\'o}z}, \&
  {Hamanowicz}}]{Soszynski:2016}
{Soszy{\'n}ski}, I., {Pawlak}, M., {Pietrukowicz}, P., {et~al.} 2016, \actaa,
  66, 405, \dodoi{10.48550/arXiv.1701.03105}

\bibitem[{{Talbot} \& {Thrane}(2018)}]{Talbot:2018}
{Talbot}, C., \& {Thrane}, E. 2018, \apj, 856, 173,
  \dodoi{10.3847/1538-4357/aab34c}

\bibitem[{{Tokovinin}(2023)}]{Tokovinin:2023}
{Tokovinin}, A. 2023, \aj, 165, 180, \dodoi{10.3847/1538-3881/acc464}

\bibitem[{{van de Kamp}(1975)}]{vandeKamp:1975}
{van de Kamp}, P. 1975, \araa, 13, 295,
  \dodoi{10.1146/annurev.aa.13.090175.001455}

\bibitem[{{van der Walt} {et~al.}(2011){van der Walt}, {Colbert}, \&
  {Varoquaux}}]{vdWalt:2011}
{van der Walt}, S., {Colbert}, S.~C., \& {Varoquaux}, G. 2011, Computing in
  Science \& Engineering, 13, 22, \dodoi{10.1109/MCSE.2011.37}

\bibitem[{{Virtanen} {et~al.}(2019){Virtanen}, {Gommers}, {Oliphant},
  {Haberland}, {Reddy}, {Cournapeau}, {Burovski}, {Peterson}, {Weckesser},
  {Bright}, {van der Walt}, {Brett}, {Wilson}, {Jarrod Millman}, {Mayorov},
  {Nelson}, {Jones}, {Kern}, {Larson}, {Carey}, {Polat}, {Feng}, {Moore}, {Vand
  erPlas}, {Laxalde}, {Perktold}, {Cimrman}, {Henriksen}, {Quintero}, {Harris},
  {Archibald}, {Ribeiro}, {Pedregosa}, {van Mulbregt}, \&
  {Contributors}}]{Scipy:2019}
{Virtanen}, P., {Gommers}, R., {Oliphant}, T.~E., {et~al.} 2019, arXiv
  e-prints, arXiv:1907.10121.
\newblock \doarXiv{1907.10121}

\bibitem[{{Weidemann}(2000)}]{Weidemann:2000}
{Weidemann}, V. 2000, \aap, 363, 647

\bibitem[{{Wiktorowicz} {et~al.}(2020){Wiktorowicz}, {Lu}, {Wyrzykowski},
  {Zhang}, {Liu}, {Justham}, \& {Belczynski}}]{Wiktorowicz:2020}
{Wiktorowicz}, G., {Lu}, Y., {Wyrzykowski}, {\L}., {et~al.} 2020, \apj, 905,
  134, \dodoi{10.3847/1538-4357/abc699}

\bibitem[{{Wu} {et~al.}(2025){Wu}, {Hadden}, {Dewberry}, {El-Badry}, \&
  {Matzner}}]{Wu:2025}
{Wu}, Y., {Hadden}, S., {Dewberry}, J., {El-Badry}, K., \& {Matzner}, C.~D.
  2025, \apjl, 982, L34, \dodoi{10.3847/2041-8213/adb751}

\bibitem[{{Yamaguchi} {et~al.}(2018){Yamaguchi}, {Kawanaka}, {Bulik}, \&
  {Piran}}]{Yamaguchi:2018}
{Yamaguchi}, M.~S., {Kawanaka}, N., {Bulik}, T., \& {Piran}, T. 2018, \apj,
  861, 21, \dodoi{10.3847/1538-4357/aac5ec}

\bibitem[{{Yamaguchi} {et~al.}(2024){Yamaguchi}, {El-Badry}, {Rees}, {Shahaf},
  {Mazeh}, \& {Andrae}}]{Yamaguchi:2024}
{Yamaguchi}, N., {El-Badry}, K., {Rees}, N.~R., {et~al.} 2024, \pasp, 136,
  084202, \dodoi{10.1088/1538-3873/ad6809}

\bibitem[{{Zonca} {et~al.}(2019){Zonca}, {Singer}, {Lenz}, {Reinecke},
  {Rosset}, {Hivon}, \& {Gorski}}]{Zonca:2019}
{Zonca}, A., {Singer}, L., {Lenz}, D., {et~al.} 2019, The Journal of Open
  Source Software, 4, 1298, \dodoi{10.21105/joss.01298}

\end{thebibliography}
\bibliographystyle{aasjournal}

\end{document}